\documentclass[11pt,preprint]{aastex}
\usepackage{mycommands}
\usepackage{verbatim}
\usepackage{graphicx}
\usepackage[percent]{overpic}
\usepackage{subfigure}
\usepackage{esint}
\usepackage{url}
\newcommand{\ol}{\overline}
\newcommand{\olM}{\ol{\mathcal{M}}_{\rm w}}
\begin{document}


\title{Incompressible Modes Excited by Supersonic Shear in Boundary Layers: Acoustic CFS Instability}

\author{Mikhail A. Belyaev}
\affil{Astronomy Department, University of California, Berkeley, CA 94720 \\ mbelyaev@berkeley.edu}

\begin{abstract}
We present an instability for exciting incompressible modes (e.g. gravity or Rossby modes) at the surface of a star accreting through a boundary layer. The instability excites a stellar mode by sourcing an acoustic wave in the disk at the boundary layer, which carries a flux of energy and angular momentum with the opposite sign as the energy and angular momentum density of the stellar mode. We call this instability the acoustic CFS instability, because of the direct analogy to the Chandrasekhar-Friedman-Schutz instability for exciting modes on a rotating star by emission of energy in the form of gravitational waves. However, the acoustic CFS instability differs from its gravitational wave counterpart in that the fluid medium in which the acoustic wave propagates (i.e.\ the accretion disk) typically rotates faster than the star in which the incompressible mode is sourced. For this reason, the instability can operate even for a non-rotating star in the presence of an accretion disk. We discuss applications of our results to high-frequency quasi-periodic oscillations in accreting black hole and neutron star systems and dwarf nova oscillations in cataclysmic variables.
\end{abstract}

\section{Introduction}

Accretion is a mass-transfer process occurring in a wide variety of astrophysical systems. We focus on the specific case of thin disk accretion onto an unmagnetized central object, or a magnetized object for which the Alfv\'en radius of the accretion flow lies inside the inner edge of the disk. Unless the accretor is a black hole, material accretes through a boundary layer at the surface of the central object, where it adjoins the disk. An alternative to boundary layer accretion is magnetospheric accretion, which occurs when the magnetic field is strong enough to disrupt the disk and channel accretion onto the magnetic poles \citep{GhoshLamb,Koldobaetal}

Observational evidence for boundary layer accretion can be found in cataclysmic variable (CV) systems that undergo changes in the accretion rate by orders of magnitude between outburst and quiescence \citep{DN_Instability}. \cite{WMM03} present observations that span the entire outburst cycle in the CV SS Cygni from quiescence through outburst and back to quiescence. They observe a lag in the UV and X-ray compared to the optical, which is associated with the time it takes for the disk to flow from the instability radius to the stellar surface \citep{LivioPringle}. For most of the outburst duration, an optically thick boundary layer that radiates in EUV is present \citep{Pringle1,PattersonRaymond2,PophamNarayan}, which is not seen in quiescence. In a different observation, \cite{Mukai} performed X-ray observations of the non-magnetic eclipsing CV binary HT Cassiopeia during quiescence. By modeling the eclipse shape, they found that the X-ray emitting region was $\lesssim 1.15$ the white dwarf radius, meaning the high-energy emission is spatially localized near the surface and comes from an optically thin boundary layer \citep{PringleSavonije,PattersonRaymond1}.

One of the hallmarks of boundary layer accretion is the intense shear present in the boundary layer. For instance, the dynamical width of the boundary layer scales as $\delta_{BL} \propto \mathcal{M}^{-2} R_*$ \citep{Pringle1,BRS}. Here the Mach number of the flow is given by $\mathcal{M} = V_K(R_*)/c_s$, where $V_K(R_*)$ is the Keplerian velocity at the surface of the star and $c_s$ is the sound speed in the boundary layer. As an example, a white dwarf in outburst has $\mathcal{M} \sim 40$, so the boundary layer is radially thin compared to the stellar radius. At the same time, for a white dwarf rotating below breakup, as much energy must be dissipated in the boundary layer as in the entire disk\footnote{This assumes a Keplerian disk which has gravitational potential and kinetic energy in the ratio ${\rm KE} = -{\rm PE}/2.$}. Thus, the shear present in the boundary layer provides a tremendous reservoir of free energy for exciting fluid instabilities.

\cite{BR} showed analytically that a radially thin boundary layer ($\delta_{BL} \ll R_*$) across which the drop in the azimuthal velocity is supersonic ($\Delta v_\phi/c_s \gtrsim 1$) is unstable to supersonic shear instabilities. These instabilities are the plane-parallel cousins of the Papaloizou-Pringle class of instabilities \citep{PP,NGG,Glatzel} and differ markedly from the Kelvin-Helmholtz instability in that they excite acoustic waves. \cite{BR,BRS1,BRS2,HertfelderKley} showed using simulations that acoustic waves excited by these supersonic shear instabilities are effective at transporting angular momentum in the boundary layer. 

The supersonic shear instability mechanism is expected on analytical grounds and is robustly observed in simulations of the boundary layer. However, previous studies focused exclusively on excitation of acoustic waves (p-modes) in both the disk and the star. In this paper we propose a mechanism by which a supersonic shear flow over the surface of a central object excites incompressible stellar modes by emission of acoustic radiation into the accretion disk. These incompressible stellar modes are ones for which pressure is not the dominant restoring force, such as internal gravity modes (g-modes) and Rossby modes (r-modes). We call our mechanism the ``acoustic CFS instability", because it is an acoustic analog to the Chandrasekhar-Friedman-Schutz instability \citep{CFS1,CFS2}, but with acoustic waves playing the role of gravitational waves in transporting away energy and angular momentum from the system.

The paper is structured as follows. In \S \ref{governeqs} we present the governing fluid dynamical equations that we shall use to elucidate the acoustic CFS instability, and in \S \ref{modelsec} we describe the setup for the model problem we consider. In \S \ref{surfgravsec} and \S \ref{rossbysec}, we derive dispersion relations for incompressible plane-parallel surface gravity waves and interfacial shear Rossby waves, respectively. In \S \ref{stardisksec} we present computational results showing excitation of incompressible modes in hydrodynamical simulations. In \S \ref{discussion} we discuss the astrophysical importance of our results and how they can be relevant for high-frequency quasi-periodic oscillations (QPOs) in neutron star and black hole systems and dwarf nova oscillations (DNOs) in CVs. We also compare the acoustic CFS instability to other related instabilities that exist in the astrophysical literature.

\section{General Framework}
\label{governeqs}
We now describe the framework that we shall use in our stability analysis to demonstrate the acoustic CFS mechanism. For simplicity, we assume that gravity points in the cylindrical radial direction. For a spherically-symmetric gravitational field, this implies we are working in a patch confined to the equatorial midplane in $z$ and are ignoring the $z$-directed component of gravity. 

We also assume an unmagnetized flow. However, we expect our conclusions to also apply in the presence of subsonic turbulence, such as that generated by magnetorotational instability (MRI). For subsonic turbulence, $v_\text{turb} \ll c_s$, where $v_\text{turb}$ is the characteristic velocity of the turbulent eddies. The Alfv\`en velocity associated to the turbulent magnetic field is $v_A/c_s \sim v_\text{turb}/c_s \ll 1$, and the dispersion relation of sound waves is not significantly modified by the presence of a magnetic field. However, turbulence could potentially provide a source of scattering and dissipation, which we do not explicitly consider in our analytical treatment. 

On the other hand, \cite{BRS2} found using simulations that acoustic waves excited in the boundary layer can still maintain their structure even in the presence of MHD turbulence in the accretion disk. Additionally, \cite{Pessah} found that MRI field advected into the boundary layer underwent a transient swing amplification of order unity but was not amplified to equipartition. Thus, magnetic field due to the MRI advected into the boundary layer is not likely to play a dominant role for the physics of the boundary layer, which is also supported by the simulations of \cite{BRS2}. Note that the boundary layer itself has a rising rotation profile ($d\Omega/dR > 0$) and hence is linearly stable to the MRI instability.

\subsection{Governing Equations}

We start from the Euler equations in cylindrical geometry:
\begin{align}
\label{Eulerstart}
-\frac{\partial \rho}{\partial t} &=\frac{1}{R}
\frac{\partial}{\partial R}(\rho R u) +
\frac{1}{R}\frac{\partial}{\partial \theta}(\rho v) +
\frac{\partial}{\partial z} (\rho w) \\
\frac{D u}{D t} &= -\frac{1}{\rho}\frac{\partial
  P}{\partial R} + \frac{v^2}{R} - g \\
\frac{D v}{D t} &= -\frac{1}{\rho R} \frac{\partial
  P}{\partial \theta} - \frac{u v}{R} \\
\frac{D w}{D t} &= -\frac{1}{\rho}\frac{\partial P}{\partial z}
\label{Eulerend}
\end{align}
The velocities in the $(R,\phi,z)$ directions are denoted by $(u,v,w)$, $g(R)$ denotes the gravitational acceleration, and $\gamma$ is the adiabatic constant. The Lagrangian derivative is given by
\begin{align}
\frac{D}{D t} &\equiv \frac{\partial}{\partial t} + u
\frac{\partial}{\partial R} + \frac{v}{R} \frac{\partial}{\partial
  \theta} + w \frac{\partial}{\partial z}.
\end{align}

Additionally, we assume an adiabatic equation of state
\begin{align}
\label{Euleradiabatic}
\frac{D \left(P\rho^{-\gamma}\right)}{D t}  &= 0.
\end{align}
The assumption of adiabaticity is valid in the absence of heating and cooling. In practice, the instabilities that we describe naturally lead to weak oblique shocks in simulations. Although, strictly speaking, the presence of such shocks violates the adiabatic assumption, we shall invoke it for simplicity.

Next, we introduce linear perturbations to the physical quantities. The linear quantities are identified by a leading $\delta$ and we take them to be in the functional form $\delta f \propto f_R(R)
\exp[i(m \theta + k_z z - \omega t)]$, which corresponds to performing a Fourier transform in the $\phi$ and $z$ dimensions and also in time. \cite{BR} have shown that the linearized first order equations for the perturbed quantities deriving from the set of equations (\ref{Eulerstart}-\ref{Eulerend}) and (\ref{Euleradiabatic}) can be reduced to two first order coupled differential equations in $\delta P$ and $\delta u$:
\begin{align}
\label{deltaPeq}
i \left(\frac{\ol{\omega}^2}{c_s^2} - k_z^2 - \frac{m^2}{R^2}\right) \delta P &=
\rho \left(C_L \ol{\omega} + \frac{2Bm}{R} \right) \delta u
+ \frac{\ol{\omega}}{R} (R \rho \delta u)' \\
\label{deltaueq}
i\rho (\ol{\omega}^2 - g_{\rm eff} C_L - \kappa^2) \delta u &= \ol{\omega}
\delta P' + \frac{ \ol{\omega} {g_{\rm eff}}}{c_s^2} \delta P - \frac{2 \Omega
  m}{R} \delta P.
\end{align}
Here, the primes denote differentiation with respect to $R$; the frequency of the mode in the frame comoving with the fluid is defined as $\ol{\omega} \equiv \omega - m \Omega$; the sound speed is given by $\gamma P/\rho = c_s^2$; the effective gravity is given by $g_{\rm eff}(R) \equiv g(R) - \Omega^2 R$, where $\Omega(R) = V_\phi(R)/R$ is the angular frequency; Oort's constant is given by $B= (\Omega + d(R \Omega)/dR)/2$; the epicyclic frequency can be expressed as $\kappa^2 = 4 B \Omega$; the Ledoux discriminant, which is related to the buoyancy of the fluid, is given by $C_L = \gamma^{-1} d \ln P/ d R - d \ln\rho/d R$.

We simplify the problem further by confining ourselves to a local patch within the cylindrical coordinate system and using Cartesian $(x,y,z)$ coordinates within the local patch in place of $(R,\phi,z)$. In this case, we can ignore coordinate curvature terms and our linear perturbed quantities now have the form
\begin{align}
\label{perturbform}
\delta f \propto f_x(x)
\exp[i(k_y y + k_z z - \omega t)],
\end{align}
where $k_y \equiv m/R$ and 
\begin{align}
\label{olo}
\ol{\omega} = \omega - k_yV_y(x).
\end{align}
Equations (\ref{deltaPeq}) and (\ref{deltaueq}) also simplify to
\begin{align}
\label{deltaPeq1}
i \left(\frac{\ol{\omega}^2}{c_s^2} - k_z^2 - k_y^2\right) \delta P &=
\rho \left(C_L \ol{\omega} + 2Bk_y \right) \delta u
+ \ol{\omega} (\rho \delta u)' \\
\label{deltaueq1}
i\rho (\ol{\omega}^2 - g_{\rm eff} C_L - \kappa^2) \delta u &= \ol{\omega}
\delta P' + \frac{ \ol{\omega} {g_{\rm eff}}}{c_s^2} \delta P - 2 \Omega k_y \delta P,
\end{align}
where the primes now denote differentiation with respect to $x$.

\subsection{Boundary Conditions}
\label{BCsec}
We shall consider velocity and density profiles with a discontinuity that defines an interface separating two fluids of different properties. We define the fluid element displacement, $\delta \xi$, and the Lagrangian pressure perturbation $\Delta P$ via
\begin{align}
\label{xidef}
\frac{D\delta \xi}{Dt}  &= \delta u \\
\label{DPdef}
\Delta P & = \delta P + \delta \xi \frac{dP}{dx}.
\end{align}
The boundary conditions we impose at the interface between the two fluids are continuity of the fluid displacement and the Lagrangian pressure perturbation:
\begin{align}
\label{BCxi}
\delta \xi_+ &= \delta \xi_- \\
\label{BCP}
\Delta P_+  &= \Delta P_- .
\end{align}
Here the positive/negative signs denote evaluation of a perturbed quantity just above/below the interface. Equation (\ref{BCxi}) ensures that the two fluids stay in contact across the interface, and equation (\ref{BCP}) ensures that there are no infinite accelerations due to a discontinuous pressure gradient across the interface.

Using hydrostatic equilibrium, $dP/dx = -g_{\rm eff} \rho$, together with equations (\ref{xidef}) and (\ref{DPdef}), and equations (\ref{perturbform}) and (\ref{olo}), we can re-express the boundary conditions (\ref{BCxi}) and (\ref{BCP}) as
\begin{align}
\label{BCxi1}
\delta \xi &= \frac{\delta u_+}{\ol{\omega}_+} = \frac{\delta u_-}{\ol{\omega}_-} \\
\label{BCP1}
\delta P_+   &= \delta P_- +  (\rho_+ g_{\rm eff,+} - \rho_- g_{\rm eff,-}) \delta \xi.
\end{align}
We have omitted the subscript on $\delta \xi$, since $\delta \xi_+ = \delta \xi_-$.

\section{Model Problem}
\label{modelsec}
We now define a model problem setup that we shall use to illustrate the acoustic CFS instability with a concrete example. We consider two fluids with different properties separated by an interface at $x=0$. The ``upper fluid" occupies the half-plane $x>0$, and the ``lower fluid" occupies the half-plane $x<0$. We shall denote quantities evaluated in the upper/lower fluid with positive/negative subscripts. It may be useful to think of the upper fluid as being a highly-simplified model for an accretion disk, and the lower fluid as being a highly-simplified model for a star. 

The lower fluid is assumed to incompressible, which implies $c_{s,-} \rightarrow \infty$ for the sound speed in the lower fluid. By making this assumption we are filtering out high-frequency acoustic modes in the lower fluid, so that only low-frequency modes survive. On the other hand, the upper fluid is compressible with sound speed $c_s$; we shall omit the ``+" subscript for brevity, when referring to the sound speed in the upper fluid, since the sound speed is infinite in the lower fluid, so there can be no confusion. Also, we allow for a tangential velocity discontinuity between the upper and lower fluids. The exact specification of the velocity profile is left to later sections, when we isolate various modes of the model system to determine their dispersion relation.

We now make three additional assumptions to simplify the problem further. First, we assume that the length scale of the density variation in both the upper and lower fluids is much greater than the length scale variation of the mode in the $x$-direction. Thus, we take the upper and lower fluids to have constant densities, $\rho_+$ and $\rho_-$, which are not necessarily the same. Second, we assume that $C_L = 0$ in both the upper and lower fluids. In so doing, we are effectively filtering out internal gravity waves in both the upper and lower fluids. However, surface gravity waves due to the discontinuity in density at the interface and Rossby waves due to rotation or shear in the velocity profile profile are still permitted. Third, we assume two-dimensional perturbations so that $k_z = 0$. 

\subsection{Upper Fluid}

We now consider linear perturbations in the upper fluid. With the above assumptions, equations (\ref{deltaPeq1}) and (\ref{deltaueq1}) in the upper fluid simplify to 
\begin{align}
\label{deltaPeq_+}
i \left(\frac{\ol{\omega}^2}{c_s^2} - k_y^2\right) \delta P &=
 2 B\rho k_y \delta u
+ \ol{\omega} \rho \delta u' \\
\label{deltaueq_+}
i\rho (\ol{\omega}^2  - \kappa^2) \delta u &= \ol{\omega}
\delta P'  - 2 \Omega k_y \delta P.
\end{align}
However, even these equations are still rather messy to work with. To simplify them further, we shall consider two separate approximations in the upper fluid, which capture different limits of the equations.

\subsubsection{Tight-Winding and WKB Approximation}
The first approximation we consider is the tight-winding limit: $|\delta u'/k_y \delta u| \gg 1$. We also assume $B \sim \Omega$, as would be the case in an accretion disk, for example. This allows us to simplify equations (\ref{deltaPeq_+}) and (\ref{deltaueq_+}) further 
\begin{align}
\label{deltaPeq_+1}
i \left(\frac{\ol{\omega}^2}{c_s^2} - k_y^2\right) \delta P &=
\ol{\omega} \rho \delta u' \\
\label{deltaueq_+1}
i\rho (\ol{\omega}^2  - \kappa^2) \delta u &= \ol{\omega}
\delta P' .
\end{align}
Combining the two equations we find
\begin{align}
\label{pre_WKB_disrel}
\left( \frac{\ol{\omega}^2 - \kappa^2}{c_s^2} \right) \delta u + \delta u'' = 0.
\end{align}

Invoking the WKB approximation, we Fourier transform in the $x$-direction so our perturbation for $\delta u$ in the upper fluid has the functional form
\begin{align}
\label{perturbform_+}
\delta u_+ \propto \exp[i(k_x x + k_y y- \omega t)],
\end{align}
where $k_x(x)$ is given by equation (\ref{pre_WKB_disrel}) via
\begin{align}
\label{kx_WKB}
 k_x^2 =  \frac{\ol{\omega}^2 - \kappa^2}{c_s^2}  \ \ \ (\text{tight winding})
\end{align}
This defines the tight-winding dispersion relation for spiral density waves in the upper fluid \citep{GT78,BinneyTremaine}. We can put it into a more familiar form by substituting $i k_x \rightarrow i k_R \equiv \partial \delta f_-/\partial R$ and using $\ol{\omega} = \omega - m \Omega = m(\Omega_P-\Omega)$:
\begin{align}
\label{spiraldensity}
m^2(\Omega-\Omega_P)^2 = c_s^2 k_R^2 + \kappa^2.
\end{align}

\subsubsection{Plane-Parallel Approximation: $\ol{\omega} \gg \Omega$}
\label{planeparallelsec}
Another approximation that we shall find particularly useful during the course of our linear analysis is $\ol{\omega} \gg \Omega$, and for reasons that will become clear shortly, we refer to this as the plane-parallel approximation. In the limit $\ol{\omega} \gg \Omega$, equations (\ref{deltaPeq_+}) and (\ref{deltaueq_+}) reduce to 
\begin{align}
\label{deltaPeq_+_cart}
i \left(\frac{\ol{\omega}^2}{c_s^2} - k_y^2\right) \delta P &=
\rho k_y  V_y' \delta u
+ \ol{\omega} \rho \delta u' \\
\label{deltaueq_+_cart}
i\rho \ol{\omega} \delta u &= \delta P'
\end{align}
with the low-frequency terms containing $\Omega$ having dropped out.

We can put equations (\ref{deltaPeq_+_cart}) and (\ref{deltaueq_+_cart}) into a cleaner form by defining
\begin{align}
\label{kx_cart}
k_x^2 &\equiv  \ol{\omega}^2/c_s^2 - k_y^2 \ \ \ (\text{plane-parallel}) \\
\label{phidef}
\delta \varphi &\equiv \delta u/k_x.
\end{align}
Note that the definitions for $k_x$ are different in the tight-winding/WKB case (equation (\ref{kx_WKB})) compared to the plane-parallel case (equation (\ref{kx_cart})), which is due to differences in the approximations we have in the two cases. Using equations (\ref{kx_cart}) and (\ref{phidef}), we can write 
\begin{align}
\label{deltaPeq_+_cart1}
i  \delta P &=
- \rho \left(\frac{\ol{\omega}}{k_x}\right)' \delta \varphi
+ \rho \left(\frac{\ol{\omega}}{k_x}\right)  \delta \varphi' \\
\label{deltaueq_+_cart1}
i k_x^2 \left(\frac{\ol{\omega}}{k_x}\right) \delta \varphi &= \delta P'.
\end{align}
We can combine equations (\ref{deltaPeq_+_cart1}) and (\ref{deltaueq_+_cart1}) into a single second order differential equation
\begin{align}
\label{Rayleigh_eq}
\delta \varphi'' + \left(k_x^2 - \frac{(\ol{\omega}/k_x)''}{\ol{\omega}/k_x}\right) \delta \varphi = 0. 
\end{align}

Equation (\ref{Rayleigh_eq}) is a type of modified Rayleigh equation that also applies for plane-parallel flow when buoyancy and stratification are not important \citep{Chimonas,Alexakis,BR}. Thus, in the limit $\ol{\omega} \gg \Omega$, the linear stability problem reduces to the plane-parallel case. As a point of note, we mention that the assumption of $k_z = 0$ is not strictly necessary in the limit $\ol{\omega} \gg \Omega$. This is because for a plane-parallel shear flow, Squire's theorem states that any three-dimensional perturbation is equivalent to a two-dimensional perturbation with wavevector along the flow, but having a reduced velocity \citep{Howard,BR}.

When there is no shear in the upper fluid $(V_y' = 0)$, $k_x$ is constant and $\ol{\omega}' = -k_y V_y' = 0$, so we can reduce equation (\ref{Rayleigh_eq}) further to the form
\begin{align}
\delta u'' + k_x^2 \delta u = 0, \ \ \ \text{($V_y' = 0$, no shear)}.
\end{align}
Thus, perturbations for $\delta u$ in the upper fluid again have the functional form (\ref{perturbform_+}) with $k_x$ now given by equation (\ref{kx_cart}). It then follows that the perturbations in the upper fluid correspond to plane-parallel sound waves when $k_x^2 > 0$. When $k_x^2 < 0$ the upper fluid perturbations are evanescent pressure perturbations, or edge waves, that decay away from the interface between the two fluids as $k_y x \rightarrow \infty$.

\subsection{Lower Fluid}
We now consider the propagation of linear disturbances in the lower fluid, and we can again start from equations (\ref{deltaPeq_+}) and (\ref{deltaueq_+}). However, because the lower fluid is incompressible, it has $c_{s,-} \rightarrow \infty$, and hence equations (\ref{deltaPeq_+}) and (\ref{deltaueq_+}) immediately simplify to
\begin{align}
\label{deltaPeq_-}
-i k_y^2 \delta P &=
 2 B\rho k_y \delta u
+ \ol{\omega} \rho \delta u' \\
\label{deltaueq_-}
i\rho (\ol{\omega}^2  - \kappa^2) \delta u &= \ol{\omega}
\delta P'  - 2 \Omega k_y \delta P.
\end{align}
Combining equations (\ref{deltaPeq_-}) and (\ref{deltaueq_-}), we find after cancellation of terms the following second-order differential equation:
\begin{align}
\label{incompress_diffeq}
\ol{\omega}^2 \left(k_y^2 \delta u - \delta u'' \right) - 2\ol{\omega} k_y B' \delta u = 0,
\end{align}
where $2B' = (\Omega' + V_y'')$.

From now on we shall focus on the special case when $B'=0$, which applies in two important limiting cases. The first is the case of uniform rotation with constant $\Omega$. The second is in the plane-parallel limit, $\ol{\omega} \gg \Omega$, with a constant linear shear, $S \equiv V_y'$. Note that there is no assumption on the magnitude of the shear in the plane-parallel limit.

When $B'=0$, it follows from equation (\ref{incompress_diffeq}) that perturbations for $\delta u$ in the lower fluid have the form
\begin{align}
\label{perturbform_-}
\delta u_- \propto \exp[k_y x + i(k_y y- \omega t)],
\end{align}
where we have chosen the sign such that the perturbation vanishes as $x \rightarrow -\infty$. Thus, the perturbation in the lower fluid can only exist as an edge wave at the interface in the lower fluid if $B' = 0$. This is not surprising, since we have filtered out internal Rossby modes by choosing $B'=0$, we have already filtered out internal gravity modes by neglecting buoyancy ($C_L = 0$), and also we have filtered out acoustic modes by making the lower fluid incompressible. Thus, only interfacial edge modes can exist in the lower fluid within our model setup.

\section{Surface Gravity Wave Excited by Acoustic CFS Instability}
\label{surfgravsec}

We now solve for the dispersion relation when the velocity profile is given by
\begin{align}
V_y(x) &= \left\{
   \begin{array}{lr}
      0 & x < 0 \\
      V_+ & x > 0
   \end{array} \right. ,
\label{shear_prof_ros}
\end{align}
and there is a constant gravitational acceleration $g$ which is directed in the $-\xhat$ direction. The densities of the upper and lower fluids are given by $\rho_+$ and $\rho_-$, respectively, where $\rho_+ < \rho_-$ so the system is stably-stratified.

If the upper fluid is absent ($\rho_+ = 0$), then the setup describes a surface gravity wave. When the upper fluid is present but $V_+ \ll c_s$, then our setup describes the classical Kelvin-Helmholtz (KH) setup with gravity. However, we shall show that when the surface gravity wave propagates supersonically with respect to the upper fluid, a sound wave is excited in the upper fluid. This sound wave carries away energy and angular momentum, leading to either damping or excitation of the surface gravity wave via emission of acoustic radiation.  
\subsection{Dispersion Relation}
\label{disrel_sec}
We now derive the dispersion relation of the system when the upper fluid is compressible, and the phase speed of the surface wave with respect to the upper fluid is arbitrary compared to $c_s$. We work in the plane parallel approximation described in \S \ref{planeparallelsec}. To simplify the notation, we shall define
\ba
\label{olomega}
\ol{\omega} \equiv \omega - k_y V_+.
\ea
Note that in the notation of \S \ref{BCsec}, $\ol{\omega}_+ = \ol{\omega}$. Also, since the lower fluid is at rest, $\ol{\omega}_- = \omega$. With this notation, the boundary conditions at the interface, equations (\ref{BCxi1}) and (\ref{BCP1}), become
\begin{align}
\label{dxigrav}
\delta \xi &= \frac{\delta u_+}{\ol{\omega}} = \frac{\delta u_-}{\omega} \\
\label{dPgrav}
\delta P_+   &= \delta P_- +  g(\rho_+ - \rho_-) \delta \xi,
\end{align}
where we have used the fact that in the plane parallel approximation $g_{\rm eff} = g$. 

Next, we use the forms of the perturbations described by equations (\ref{perturbform_+}) and (\ref{perturbform_-}) in the upper and lower fluids respectively. Then we can use equation (\ref{deltaueq_+_cart}) to express $\delta P_+$ in terms of $\delta u_+$. Similarly we can use equation (\ref{deltaPeq_-}) with $B=0$ (since $V_y' = 0$ in the lower fluid) to express $\delta P_-$ in terms of $\delta u_-$. Defining $k_{x,+}$ via equation (\ref{kx_cart}) in the upper fluid, we can then combine equations (\ref{dxigrav}) and (\ref{dPgrav}) to form the dispersion relation
\ba
\label{disrel}
\rho_- \omega^2 -\frac{k_y}{i k_{x,+}} \rho_+ \ol{\omega}^2 - g k_y(\rho_- - \rho_+) = 0. 
\ea
If we define
\begin{align}
\label{adef}
a_+ \equiv -\frac{k_y}{i k_{x,+}}
\end{align}
and introduce the density ratio
\begin{align}
\label{epsdef}
\epsilon &\equiv \rho_+/\rho_- < 1,
\end{align}
we can solve for $\omega$ in equation (\ref{disrel}):
\ba
\label{disrel1}
\omega = \frac{k_yV_+ \epsilon a_+ \pm \sqrt{g k_y(1-\epsilon)(1+\epsilon a_+) - (k_yV_+)^2\epsilon a_+}}{1+\epsilon a_+}. 
\ea
Defining the characteristic frequency
\begin{align}
\label{V0eq}
\omega_0 \equiv \sqrt{g k_y (1-\epsilon)},
\end{align}
the dispersion relation (\ref{disrel1}) can be written as
\begin{align}
\label{disrel2}
\omega = \omega_0 \left[\frac{(k_yV_+/\omega_0) \epsilon a_+ \pm \sqrt{1+\left(1 - (k_yV_+/\omega_0)^2\right) \epsilon a_+}}{1+\epsilon a_+}\right].
\end{align}

Equation (\ref{disrel2}) is generally an implicit expression for $\omega$, which enters into $a_+$ through $k_{x,+}$. However, notice that if $a_+ = 0$, then the phase speed in the lower fluid is given by equation (\ref{V0eq}). We now examine how $a_+$ depends on the Mach number in the upper fluid. We begin by focusing on the low Mach number incompressible limit for which we recover the KH instability. Then we proceed to the high Mach number limit for which the significance of $\omega_0$ as a characteristic frequency becomes apparent.

\subsection{Incompressible Kelvin-Helmholtz Limit}
We verify equation (\ref{disrel}) in the limit where the upper fluid is incompressible, in which case it reduces to the well-known dispersion relation for the KH instability in the presence of gravity. The upper fluid is incompressible in the limit $c_s \rightarrow \infty$, in which case from equation (\ref{kx_cart}) $k_{x,+}^2 = -k_y^2$. Thus, the perturbation in the upper fluid is evanescent, and we must have $k_{x,+} = i k_y$, so that it vanishes as $x \rightarrow \infty$.  

From the definition of $a_+$ in equation (\ref{adef}), we find in the incompressible limit that $a_+ = 1$. Substituting this value into equation (\ref{disrel}) and expanding, the dispersion relation becomes
\ba
\label{disrel_incompress}
(\rho_- + \rho_+) \omega^2 -2 \rho_+ k_yV_+\omega + \rho_+(k_yV_+)^2 - g k_y(\rho_- - \rho_+) = 0.
\ea
This is indeed the well-known dispersion relation for the Kelvin-Helmholtz instability in the presence of gravity. In the absence of shear across the interface ($V_+=0$), the phase speed of the gravity wave according to equation (\ref{disrel_incompress}) is given by
\begin{align}
\label{Vgeq}
\frac{\omega}{k_y} = \sqrt{\frac{g}{k_y}\frac{1-\epsilon}{1+\epsilon}},
\end{align}
where $\epsilon$ is the density ratio defined in equation (\ref{epsdef}).

\subsection{General Properies of the Dispersion Relation}
We now relax the incompressibility assumption and define
\begin{align}
\label{machnum}
\ol{\mathcal{M}}_{\rm w}^2 \equiv  \left(\frac{V_+ -\omega/k_y}{c_s} \right)^2  =  \left(\frac{\ol{\omega}}{k_y c_s} \right)^2.
\end{align}
This is a mathematical definition, but if $\omega$ is real, then $\ol{\mathcal{M}}_{\rm w}$ can be physically interpreted as {\it the Mach number of the phase speed of the wave relative to the upper fluid}. In terms of the definition (\ref{machnum}), we can write
\begin{align}
\label{k_upper}
k_{x,+}^2 &= -k_y^2\left(1-\olM^2\right) \\
\label{a_upper}
a_+ &= \pm \left(1-\olM^2\right)^{-1/2},
\end{align}
where the choice of sign is dictated by the boundary conditions. 

It is interesting to see what happens to $k_{x,+}$ and $a_+$ in equations (\ref{k_upper}) and (\ref{a_upper}) as $\olM$ is increased from $0$ to $\infty$. For $\olM = 0$ (the incompressible case), $k_{x,+}$ is imaginary so the mode in the upper fluid is evanescent away from the interface. As we start to increase $\olM$ with $\olM < 1$, $k_{x,+}$ decreases in magnitude, so the evanescence length of the mode in the upper fluid becomes longer. As $\olM \rightarrow 1$ from below, we have $k_{x,+} \rightarrow 0$, and for $\olM = 1$ the evanescence length of the mode in the upper fluid becomes infinite!

The physics changes drastically as we cross over into the supersonic regime $\olM > 1$. The $x$-component of the wavevector, $k_{x,+}$, now has a real component, and the disturbance in the upper fluid corresponds to a sound wave propagating away from the interface. Furthermore, as $\olM \rightarrow \infty$ we see that $k_{x,+} \rightarrow \infty$, so the sound wave that appears in the upper fluid for $\olM > 1$ becomes ``tightly-wound". Moreover, for $\olM \rightarrow \infty$ we have $a_+ \rightarrow 0$, in which case the angular frequency of the mode is given by $\omega = \omega_0$. This is the physical significance of the angular frequency $\omega_0$ defined in equation (\ref{V0eq}).

\subsection{Dispersion Relation in the Hypersonic Limit}
\label{hypersonic_sec}
\label{disrelgrav}
We now discuss solutions to equation (\ref{disrel2}) in the hypersonic limit, $\olM \gg 1$. In this limit, $|a_+| \ll 1$ and since $\epsilon < 1$ this implies $|\epsilon a_+| \ll 1$. Expanding to first order in $\epsilon a_+$, we may write equation (\ref{disrel2}) as
\begin{align}
\omega &=  \pm \omega_0 \left[1 - i \frac{\epsilon }{2} \frac{k_y}{k_{x,+}} \left(\frac{k_yV_+}{\omega_0} \mp 1 \right)^2 \right],
\label{disrel3}
\end{align}
which shows that the angular frequency is real at leading order and equal to $\omega_0$. The first order correction proportional to $\epsilon a_+$ is purely imaginary to leading order. This is because $k_{x,+}$ is generally real to leading order for $\olM \gg 1$, which can be seen by substituting equation (\ref{disrel3}) into equation (\ref{k_upper}). The upper or lower set of signs in equation (\ref{disrel3}) is chosen according to whether the mode is prograde or retrograde in the frame of the lower fluid: ${\rm sgn}(\omega) \equiv {\rm sgn}({\rm Re}[\omega])$. 

It is also important to note that the sign of $a_+$ in equation (\ref{growdamp}) depends on the sign of $k_{x,+}$. This is determined according to the condition of outgoing radiation in the frame of the upper fluid, which ensures that no radiation propagates inward from $x = \infty$ towards the interface:
\begin{align}
\label{rad_out1}
\frac{{\rm Re} [\ol{\omega}]}{{\rm Re}[k_{x,+}]} > 0, \ \ \text{(outwardly propagating radiation)}.
\end{align}
Using equation (\ref{k_upper}) together with the radiation boundary condition and the definition of $a_+$ from equation (\ref{adef}), we can write for $\olM \gg 1$
\begin{align}
\label{k_hyper}
k_{x,+} &= {\rm sgn}(\ol{\omega})k_y\sqrt{\olM^2 - 1}.
\end{align}
Here, ${\rm sgn}(\ol{\omega}) \equiv {\rm sgn}({\rm Re}[\ol{\omega}])$ is positive/negative for modes that are prograde/retrograde in the frame of the upper fluid.

Substituting equation (\ref{k_hyper}) into equation (\ref{disrel3}), we can write the dispersion relation in the hypersonic regime with the boundary conditions taken into account:
\ba
\label{disrel4}
\omega = {\rm sgn}(\omega) \omega_0 \left[1 -  {\rm sgn}(\ol{\omega}) \frac{i}{2} \epsilon (\olM^2 - 1)^{-1/2} \left(\frac{k_yV_+}{{\rm sgn}(\omega) \omega_0} - 1 \right)^2 \right].
\ea
The sign of the imaginary term in the dispersion relation determines whether an instability exists. In particular, we have the following criterion for whether a mode grows or damps due to emission of acoustic radiation:
\ba
\label{growdamp}
{\rm sgn(\omega)} \times {\rm sgn(\ol{\omega})} \left\{
   \begin{array}{ll}		
      < 0,& \text{growing mode} \\
      > 0,& \text{decaying mode}
   \end{array} \right. .
\ea

We see that only modes for which ${\rm sgn(\omega)}$ and ${\rm sgn(\ol{\omega})}$ are of opposite sign are unstable. These are modes which are retrograde in the reference frame of the upper fluid, but are drawn prograde in the reference frame of the lower fluid by the relative velocity of the two fluids. These modes have the opposite sign of momentum and energy in the lower fluid as the flux of energy and momentum carried by the sound wave in the upper fluid. Hence, they can be understood in terms of the negative wave energy principle, and we provide a physical interpretation for the growth rate in the next section.

\section{Interfacial Shear Rossby Wave Sourced by Acoustic CFS Instability}
\label{rossbysec}
In \S \ref{surfgravsec}, we demonstrated that a surface gravity can be unstable to the acoustic CFS mechanism when it propagates supersonically relative to a compressible medium in the presence of shear. Here we show the same is true of an interfacial shear Rossby wave. We again work in the plane-parallel approximation of \S \ref{planeparallelsec}, which implies that the shear frequency is much greater than the rotational frequency $S \equiv V'  \gg \Omega$. This is most applicable to the case of e.g.\ an astrophysical boundary layer.

We again work in plane-parallel geometry and consider an equilibrium background state consisting of two constant density fluids separated by an interface at $x=0$.  The lower fluid occupies the half-plane $x < 0$ and is incompressible with density $\rho_-$. The upper fluid occupies the half-plane $x > 0$ and is compressible with constant sound speed $c_s$ and density $\rho_+$. As before, we denote quantities in the upper and lower fluids by the subscripts ``$+$" and ``$-$", respectively.

The velocity profile we consider is given by
\begin{align}
V_y(x) &= \left\{
   \begin{array}{lr}
      Sx & x < 0 \\
      V_+ & x > 0
   \end{array} \right. ,
\label{shear_prof_ros}
\end{align}
where $V_+$ and $S$ are constants. From equation (\ref{shear_prof_ros}) it is clear that $V_+$ defines the velocity of the upper fluid, whereas $S \equiv dV_y/ dx$ is a constant linear shear in the lower fluid. To keep things simple, there is no gravity in our model problem, and we are concerned with a pure interfacial Rossby wave sourced by shear.

\subsection{Dispersion Relation}
As in \S \ref{disrel_sec}, we define $\ol{\omega} \equiv \omega - k_y V_+$. Starting from the boundary conditions at the interface, equations (\ref{BCxi1}) and (\ref{BCP1}), we can write 
\begin{align}
\label{dxiross}
\delta \xi &= \frac{\delta u_+}{\ol{\omega}} = \frac{\delta u_-}{\omega} \\
\label{dPross}
\delta P_+   &= \delta P_-,
\end{align}
where we have used the fact that $g_{\rm eff} = 0$, since we do not have gravity in the model problem and are working in the plane-parallel approximation.

As before, we use the forms of the perturbations described by equations (\ref{perturbform_+}) and (\ref{perturbform_-}) in the upper and lower fluids respectively. Then we can use equation (\ref{deltaueq_+_cart}) to express $\delta P_+$ in terms of $\delta u_+$. Similarly we can use equation (\ref{deltaPeq_-}) with $B=S/2$ to express $\delta P_-$ in terms of $\delta u_-$. Defining $k_{x,+}$ via equation (\ref{kx_cart}) in the upper fluid, we can then combine equations (\ref{dxiross}) and (\ref{dPross}) to form the dispersion relation
\begin{align}
\label{rossbydis}
\omega(\omega + S) + \epsilon a_+  \ol{\omega}^2 = 0.
\end{align}
Here, $\epsilon$ gives the density ratio via equation (\ref{epsdef}), and $a_+$ is defined via equation (\ref{adef}). We can solve for $\omega$ in equation (\ref{rossbydis}) to yield
\begin{align}
\label{rossby_sol}
\omega = \frac{-S/2+k_yV_+ \epsilon a_+ \pm \sqrt{(S/2)^2 - \epsilon a_+k_yV_+(S+k_yV_+)}}{1+\epsilon a_+}.
\end{align}

\subsection{Incompressible Limit}
\label{rossby_incompress}
We now check our dispersion relation, equation (\ref{rossbydis}), by comparing it to known results in the limit when the upper fluid is incompressible. In this limit, $c_s \rightarrow \infty$ and $a_+ = 1$ in the upper fluid. Thus equation (\ref{rossbydis}) becomes 
\ba
\label{rossbydis_incompress}
\omega(\omega + S) + \epsilon  \ol{\omega}^2 = 0.
\ea
If $S= 0$ and there is no shear in the lower fluid, equation (\ref{rossbydis_incompress}) is the dispersion relation for KH instability in the absence of gravity.

Alternatively, we can consider the case of $\epsilon =1$ and no velocity discontinuity at the interface ($V_+ = 0$). This implies $\ol{\omega} = \omega$, in which case
\ba
\omega (\omega + S) + \omega^2 = 0.
\ea
This has the solution $\omega = -S/2$, which describes a Rossby edge wave at a discontinuity in the vorticity for a piecewise linear velocity profile \citep{Heifetz}. Notice that the Rossby wave propagates unidirectionally, which can be understood by analogy to a rotating system and conservation of vorticity. In a rotating system undergoing solid body rotation, the vorticity is given by $\varpi = 2\Omega$, whereas in the case of linear shear that we consider here, the vorticity is $\varpi = S$. In both cases, though, vorticity conservation in an incompressible flow of uniform density is the physical constraint determining the dynamics and leading to unidirectional propagation of the Rossby wave. 

\subsection{Dispersion Relation in the Hypersonic Limit}
This dispersion relation (\ref{rossby_sol}) has many parallels with the one we found for surface gravity waves. In particular, the subsonic regime, $\mathcal{M}_{\rm w} < 1$, is qualitatively similar to the incompressible case $\mathcal{M}_{\rm w} = 0$ with the disturbance in the upper fluid evanescent away from the interface. On the other hand, in the supersonic regime, $\mathcal{M}_{\rm w} > 1$, a sound wave is launched in the upper fluid that carries energy and  momentum to infinity.

In this section, we shall consider the hypersonic limit $\mathcal{M}_{\rm w} \gg 1$, and we expand equation (\ref{rossby_sol}) to first order in $\epsilon a_+$
\begin{align}
\omega &=-\left( \frac{1}{2} \mp \frac{1}{2} \right)S +  i \epsilon \frac{k_y}{k_{x,+}} \left[ \left( \frac{1}{2} \mp \frac{1}{2} \right)S + k_yV_+ \mp \frac{k_y V_+(S+k_y V_+)}{S}\right]  
\label{rossby_hyper}
\end{align}
It is useful at this point to break the solution down into $\omega_N$ corresponding to the upper set of signs in equation (\ref{rossby_hyper}) and $\omega_R$ corresponding to the lower set of signs in equation (\ref{rossby_hyper}):
\begin{align}
\label{oNeq}
\omega_N &=  - i \epsilon \frac{(k_y V_+)^2}{S} \left( \frac{k_y}{k_{x,+}} \right)\\
\label{oReq}
\omega_R &= -S + i \epsilon \frac{\left(S -  k_yV_+ \right)^2}{S} \left( \frac{k_y}{k_{x,+}} \right).
\end{align}
We have chosen this particular subscript notation, because the mode with frequency $\omega_N$ has ${\rm Re}[\omega_N] = 0$ and hence is a neutral mode that does not propagate in our chosen frame of reference. On the other hand, the mode with frequency $\omega_R$ has ${ \rm Re}[\omega_R] = -S$ and hence propagates unidirectionally in the same sense as a Rossby wave.

We must still determine the sign of $k_{x,+}$ in equations (\ref{oNeq}) and (\ref{oReq}) by imposing the radiation boundary condition in the upper fluid (equation (\ref{rad_out1})). We can then express $\omega_N$ and $\omega_R$ in terms of the wave Mach number as
\begin{align}
\label{oNeq1}
\omega_N &=  - {\rm sgn}(\ol{\omega}) i  \epsilon \frac{(k_y V_+)^2}{S} \left( \olM^2-1\right)^{-1/2}\\
\label{oReq1}
\omega_R &= -S \left[1 - {\rm sgn}(\ol{\omega}) i \epsilon \left(\frac{S -  k_yV_+}{S}\right)^2 \left(\olM^2-1\right)^{-1/2} \right].
\end{align}

\section{Star with an Accretion Disk}
\label{stardisksec}
In \S \ref{surfgravsec} and \S \ref{rossbysec} we worked under the plane-parallel assumption, which allowed us to derive simple analytical formulas in Cartesian coordinates. We now perform 2D hydrodynamical simulations in cylindrical coordinates of a Kepelarian disk adjacent to a non-rotating star. The aim of these simulations is to show that incompressible stellar modes are indeed excited and that they source acoustic waves that propagate in the disk. New physics is introduced in the star--disk setup that is not present in the plane-parallel case. In particular, the presence of an evanescent region in the disk between Lindblad resonances traps the wave, preventing it from propagating to infinity. 

\subsection{2D Hydro Star--Disk Simulation}
\label{star_disk_sec}
We perform a 2D ideal hydro simulation (no viscosity, no magnetic fields) of a star--disk system using an adiabatic equation of state with $\gamma = 5/3$. The star is initially non-rotating and the disk has a Keplerian rotation profile. The initial equilibrium state is isothermal, and both the disk and the star have an initial sound speed of $c_s= .1$ in units where $R_* = \Omega_K(R_*) = 1$. Initially, there is a discontinuity in the azimuthal velocity at $R=R_*$, which defines the separation between the disk and the star. At the start of the simulation, we introduce random small-amplitude seed perturbations in the pressure. As a result, supersonic shear instabilities are excited, which quickly smear out the velocity discontinuity into a boundary layer so the velocity profile becomes everywhere well-resolved. After an initial transient state that heats the inner part of the disk to $c_s \sim .15-.2$, the system settles into a quasi-steady state on a timescale of order ten orbits at the surface of the star.

For the simulations we use the code {\it Athena++} in cylindrical coordinates with a logarithmic radial spacing. The extent of the simulation domain is $R=[.92,13.6]$ and $\phi=[0,2\pi]$, and the resolution is $N_r \times N_\phi = 9000 \times 1536$. The need for such a high resolution in the radial direction stems from the fact that the radial pressure scale height at the surface of the star is
\ba
\label{dlnpdlnr}
\left|\frac{{\rm d} \ln P}{{\rm d} R} \right|^{-1}  = .006 R_*.
\ea
At our level of radial resolution, we resolve the radial pressure scale height, which is the smallest physical length scale in the simulation, with 20 cells. Also, we mention that even though the star extends only from the inner boundary at $R=.92$ to $R=R_* = 1$, the density at the inner boundary is approximately $6 \times 10^5$ times higher than at $R = R_*$ due to the small radial pressure scale height. The boundary conditions at both the inner and outer boundaries are ``do-nothing" boundary conditions. This means the fluid variables at the boundary take their initial values for all time. This type of boundary condition damps incident waves, unlike a reflecting boundary condition, while at the same time preventing the star from ``falling through" the inner boundary.

Given the initial isothermal temperature profile of the star\footnote{The temperature in the star remains close to its initial value of $c_s = .1$ for the duration of the simulation.}, it is easy to distinguish between high-frequency p-modes and low-frequency incompressible modes. This is because sound waves in the star do not propagate below the Brunt-V\"ais\"al\"a frequency. For an isothermal atmosphere, this frequency is given by 
\ba
\label{BVeq}
N^2 \equiv (\gamma - 1)g^2/s^2
\ea
\citep{Vallis}. Near the surface of the star, it can be expressed in terms of the Keplerian orbital frequency and the Mach number as
\ba
\label{BVstar}
N_* = \mathcal{M} (\gamma - 1)^{1/2}  \Omega_K(R_*),
\ea
where $\mathcal{M} \approx 10$ and $N_* \approx 8.2$ in code units. Modes with frequencies higher than $N_*$ are compressible p-modes, whereas those with frequencies lower than $N_*$ are the low-frequency incompressible modes that we are interested in here.

\begin{figure*}[!t]
\centering
\subfigure{\begin{overpic}
		 [width=.49\textwidth]{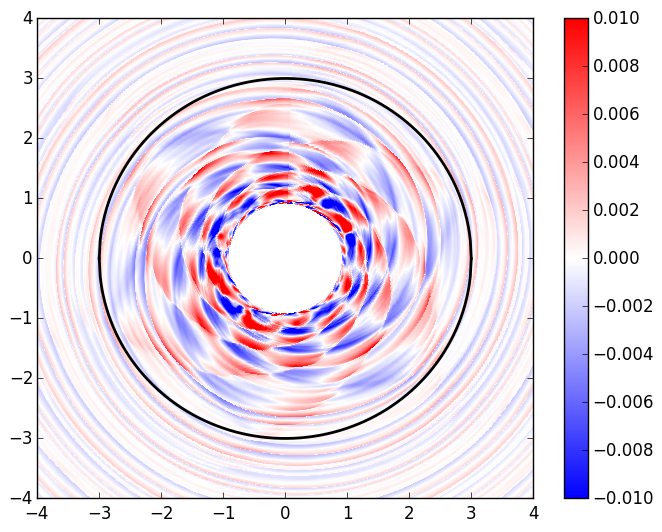}
		 \put(10,70){\Large $\rho^{1/2} v_R$}
		 \end{overpic}}
\subfigure{\begin{overpic}
		 [width=.49\textwidth]{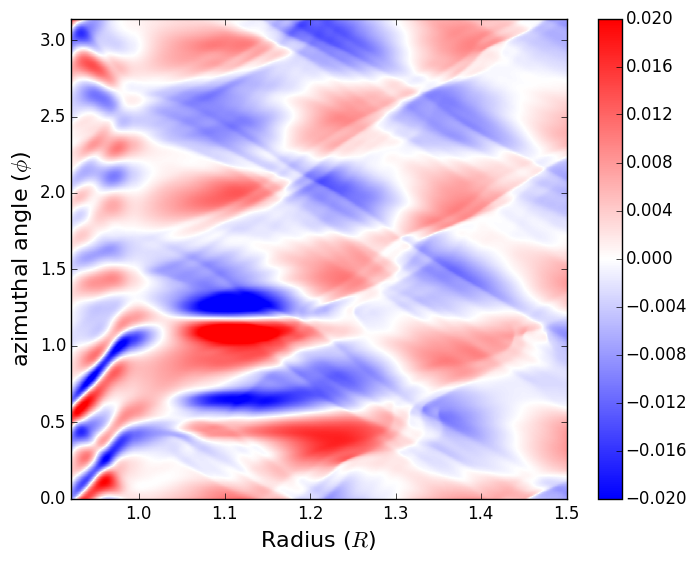}
		 \put(14,70){\Large $\rho^{1/2} v_R$}
		 \end{overpic}}
\caption{Density-weighted radial velocity, $\rho^{1/2} v_R$ for the disk--star setup. Left panel: snapshot of the disk with the ILR of the slow mode depicted as a black circle. Right panel: zoom-in to the region $R=[.92,1.5]$ and $\phi=[0,\pi]$ for the same image as the left panel.}
\label{stardiskfig}
\end{figure*}
 
Fig.\ \ref{stardiskfig} shows $\rho^{1/2}v_R$ after an elapsed time of $250$ orbital periods at the surface of the star. Because the density varies by orders of magnitude in the star, $\rho^{1/2}v_R$ is suitable for visualizing waves in the system. The left panel shows the entire star--disk system out to a radius of $R=4$. The right panel shows a zoom-in of the region $R = [.92,1.5]$ and $\phi = [0,\pi]$ with radial coordinate on the $x$-axis and azimuthal coordinate on the $y$-axis. 
 
Focusing on the right panel of Fig.\ \ref{stardiskfig}, we see that there are two modes that clearly stand out. One mode is especially prominent in the disk and has a pattern number $m = 6$. The second mode, or rather set of modes, is especially prominent inside the star ($R < 1$) and corresponds to higher values of $m$. We can determine the pattern speeds and $m$ values of the modes in the disk and in the star by Fourier transforming in time and in azimuthal angle, respectively. 

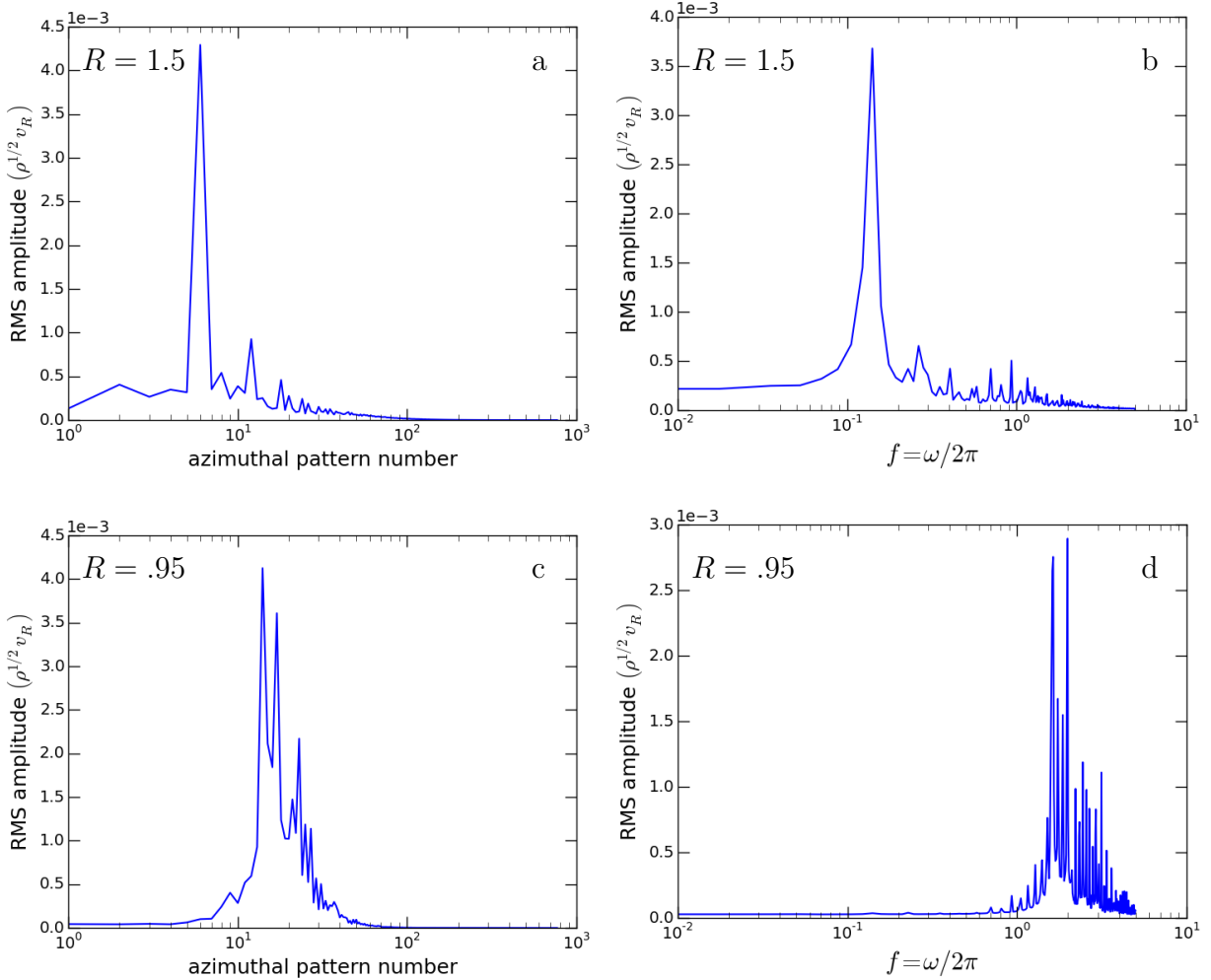
\begin{figure*}[!t]
\centering
\subfigure{\begin{overpic}
		 [width=.49\textwidth]{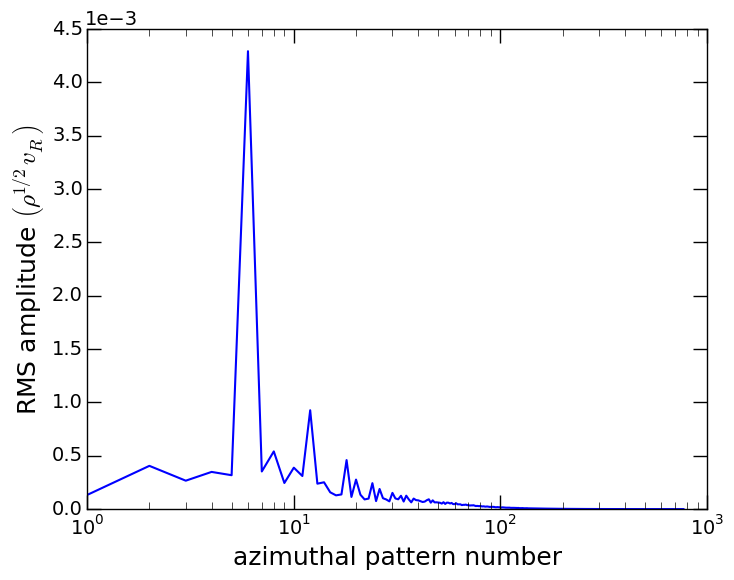}
		 \put(14,68){\large $R=1.5$}
		 \put(89,68){\large a}
		 \end{overpic}}
\subfigure{\begin{overpic}
		 [width=.49\textwidth]{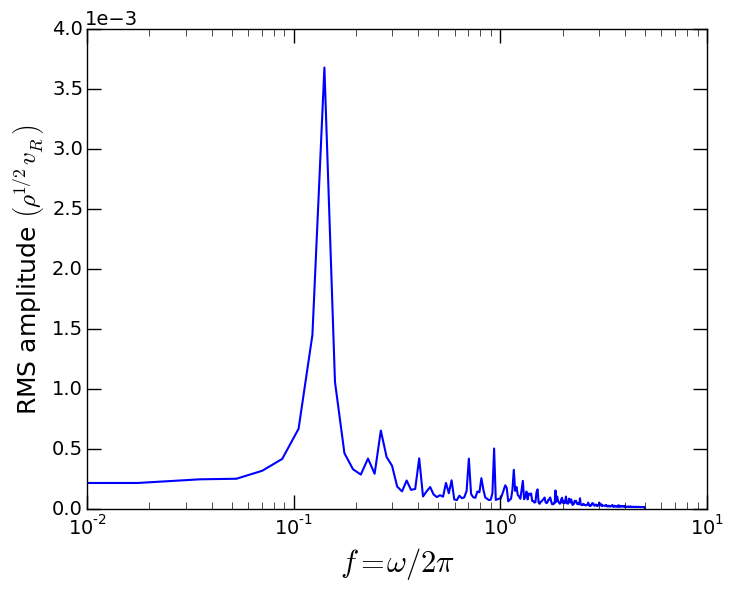}
		 \put(14,68){\large $R=1.5$}
		 \put(89,68){\large b}
		 \end{overpic}}
\subfigure{\begin{overpic}
		 [width=.49\textwidth]{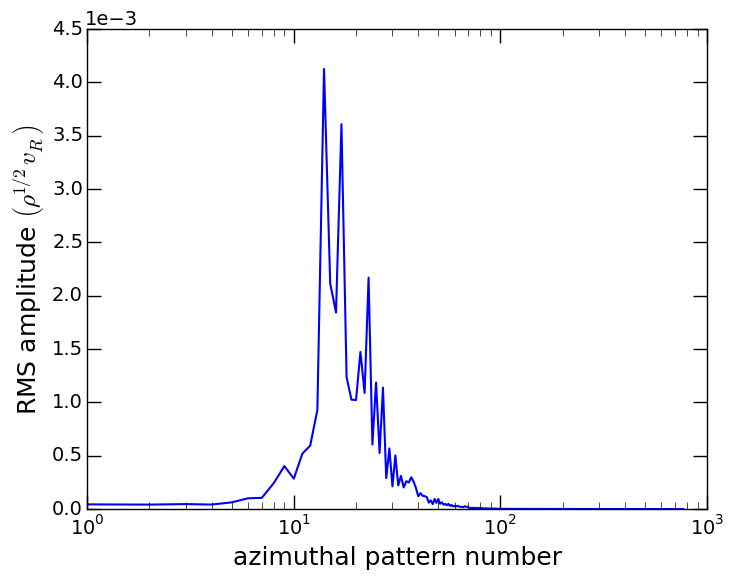}
		 \put(14,68){\large $R=.95$}
		 \put(89,68){\large c}
		 \end{overpic}}
\subfigure{\begin{overpic}
		 [width=.49\textwidth]{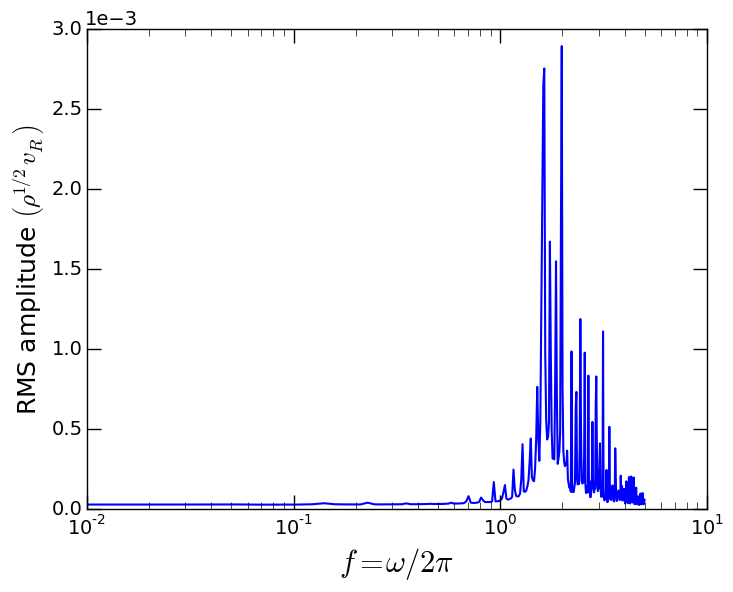}
		 \put(14,68){\large $R=.95$}
		 \put(89,68){\large d}
		 \end{overpic}}
\caption{Root mean square amplitude of $\rho^{1/2} v_R$ vs.\ pattern number (panels a and c) and frequency (panels b and d). The radius at which the power spectrum is calculated is indicated as either $R=1.5$ (inside the disk) or $R=.95$ (inside the star).}
\label{powerspectrum_fig}
\end{figure*}

Fig.\ \ref{powerspectrum_fig} shows the linear root mean square power spectrum of $\rho^{1/2}v_R$. The power spectrum is taken separately in time or in space (Fourier transform in $t$ or in $\phi$) at the radii $R = .95$ in the star (lower panels) and $R=1.5$ in the disk (upper panels). From panel a, we see that the mode in the disk indeed has a single dominant pattern number at $m=6$, and from panel b we see it has a pattern speed of $\Omega_P \approx .14-.15$. This pattern speed also agrees with the location of the Lindblad radius in the disk at $R \approx 3.1$, which gives a similar estimate of the pattern speed.

From panel c of Fig.\ \ref{powerspectrum_fig}, we see that the modes in the disk have pattern numbers in the range $m \sim 12-25$ with the dominant modes around $m \sim 13-17$. The pattern speed of the modes in the disk is in the range $\Omega_P \sim .67-.73$. Because the modes in the star have a relatively small range of pattern numbers and a high pattern speed and frequency, we shall refer to them collectively as the ``fast modes". Similarly, we shall refer to the $m=6$ mode in the disk as the ``slow mode".

Because the fast modes have $\omega = m \Omega_P \sim 10-12 > N_*$, they correspond to p-modes inside the star. In fact, their pattern speed is consistent with the ``upper branch" mode of \cite{BRS1}, which exists as an acoustic mode in both the star and the disk. The slow mode, on the other hand, has $\omega \approx .84-.9 \ll N_*$, and thus corresponds to an incompressible mode in the star coupled to an acoustic mode in the disk. This is precisely the type of mode to which our discussion of the acoustic CFS instability applies, and it is encouraging to see evidence for it in a first-principles simulation. 

Another interesting observation that can be made from the right panel of Fig.\ \ref{stardiskfig} is that the fast mode has a higher amplitude than the slow mode inside the star, even though the slow mode is more prominent than the fast mode inside of its ILR in the disk. This may suggest that the slow mode is a surface wave, the energy of which decays faster than $\rho^{-1/2}$ into the star. According to equation (\ref{Imw2}), the fastest growing modes will have the largest ratio of the wave energy and angular momentum current in the disk to the total energy and angular momentum associated with the mode in the star. Thus, it is reasonable to expect that acoustic CFS instability preferentially excites modes with energy and angular momentum concentrated at the stellar surface.

\subsection{What Sets the Pattern Number of Trapped Modes in the Disk?}
\label{pattern_number_sec}

In principle, the acoustic CFS instability should excite modes with a variety of frequencies, but a dominant low-frequency mode (the slow mode) is visible in the disk in Fig. \ref{stardiskfig}a and in the disk power spectrum in Fig. \ref{powerspectrum_fig}a,b. A natural question to ask, therefore, is what sets the pattern number of this special mode? To answer this question, one can consider the tight-winding dispersion relation for spiral density waves in the disk (equation \ref{spiraldensity}). The acoustic wave is evanescent in the region of imaginary $k_R$, and the boundaries of this region are the inner and outer Lindblad resonances, where $k_R = 0$. For a Keplerian disk, $\kappa = \Omega$, and the Lindblad resonances are located at
\begin{align}
R_{\rm LR} = \left(\frac{\Omega_P}{ 1 \pm 1/m}\right)^{-2/3}.
\end{align}

An acoustic wave propagating in the disk away from the surface of the star that encounters the inner Lindblad resonance is reflected back toward the surface of the star. If the reflected incoming wave is in phase with the outgoing excited wave, then constructive interference occurs, and the amplitude of the outgoing mode is reinforced. Thus, constructive interference occurs between the Lindblad resonance and the disk, and the amplitude of the mode is enhanced. The resonance condition for constructive interference of an acoustic mode trapped between the surface of the white dwarf and the ILR \citep{BRS} can be written as 
\begin{align}
\label{resonance_eq}
\frac{2 \pi (n+1)}{m} = 2 \Delta \phi.
\end{align}
Here, $n \ge 0$ is an integer specifying the number of nodes, and $\Delta \phi$ is the azimuthal angle traversed by a wavefront of the acoustic mode in the disk between the surface of the star and the ILR:
\begin{align}
\label{dphi_eq}
\Delta \phi &= \int_{R_*}^{R_{\rm ILR}} dR \frac{d \phi}{dR} \\
&= \int_{R_*}^{R_{\rm ILR}} dR \frac{k_R}{m}.
\end{align}

To quantify how much the magnitude of an acoustic mode in the disk near resonance is amplified due to constructive interference, we run a set of simulations with a Keplerian disk. In the simulations, the disk extends to the inner boundary of the simulation domain, $R_0$. At the inner boundary, we impose by hand a small radial velocity perturbation, $v_R \ll c_s$ that propagates with a pattern speed $\Omega_P$: 
\begin{align}
v_R(R_0) = V_{R,0} \cos(m(\phi - \Omega_P t)). 
\end{align}
In this way we are able to directly impose the value of $\Omega_P$, allowing for a controlled numerical experiment.

We use an isothermal equation of state with $c_s/V_K(R_0) = .1$, so the Mach number of the flow over the surface of the inner boundary is $\mathcal{M} = 10$. When the Mach number of the wave relative to the flow,
\ba
\olM &= (V_K(R_0) - R_0 \Omega_P)/c_s \\
&= \mathcal{M}(1 - \Omega_P/\Omega_K(R_0)),
\ea
is supersonic ($\olM > 1$), an acoustic wave is launched into the disk.  

We run two suites of 500 hydrodynamical 2D simulations extending from $R = [1,40]$ and $\phi = [0,2\pi]$ with a resolution of $N_r \times N_\phi = 500 \times 256$. Each simulation is run for a duration of $\Delta t = 3000 \Omega_K^{-1}(R_0)$. The Keplerian velocity of the disk at the inner boundary and the radius of the inner boundary are normalized to one ($\Omega_K(R_0) = 1$ and $R_0 = 1$), which sets the units for the dimensions of space and time. The pattern number of the mode used in the simulations is $m=2$. 

For each suite of simulations, we use logarithmically spaced intervals in $\Omega_P$ in the range $\Omega_P = [.01,.5]$. There is one value of $\Omega_P$ per simulation, and this is the only parameter which is varied between simulations within a given suite. For the value $\Omega_P = .5$, the ILR is located at $R = 1$ and is adjacent to the inner boundary. To study the saturation amplitude of the trapped modes near resonance, one suite of simulations uses $V_{R,0} = .01$ and the other suite uses $V_{R,0} = .003$ for the amplitude of the imposed perturbation at the inner boundary. This is the only parameter which is varied between the two suites of simulations. 

\begin{figure*}[!b]
\centering
\subfigure{\includegraphics[width=.49\textwidth]{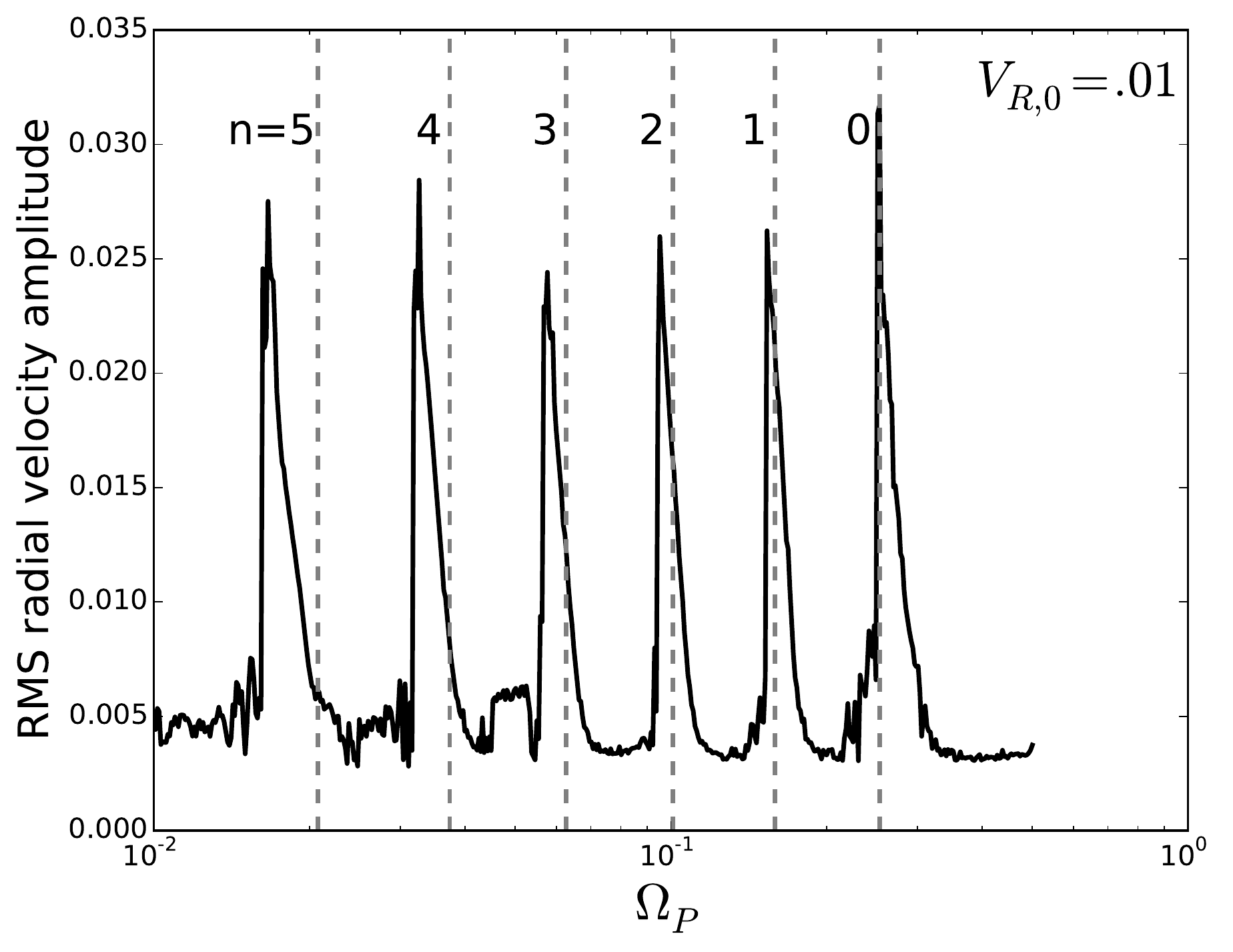}}
\subfigure{\includegraphics[width=.49\textwidth]{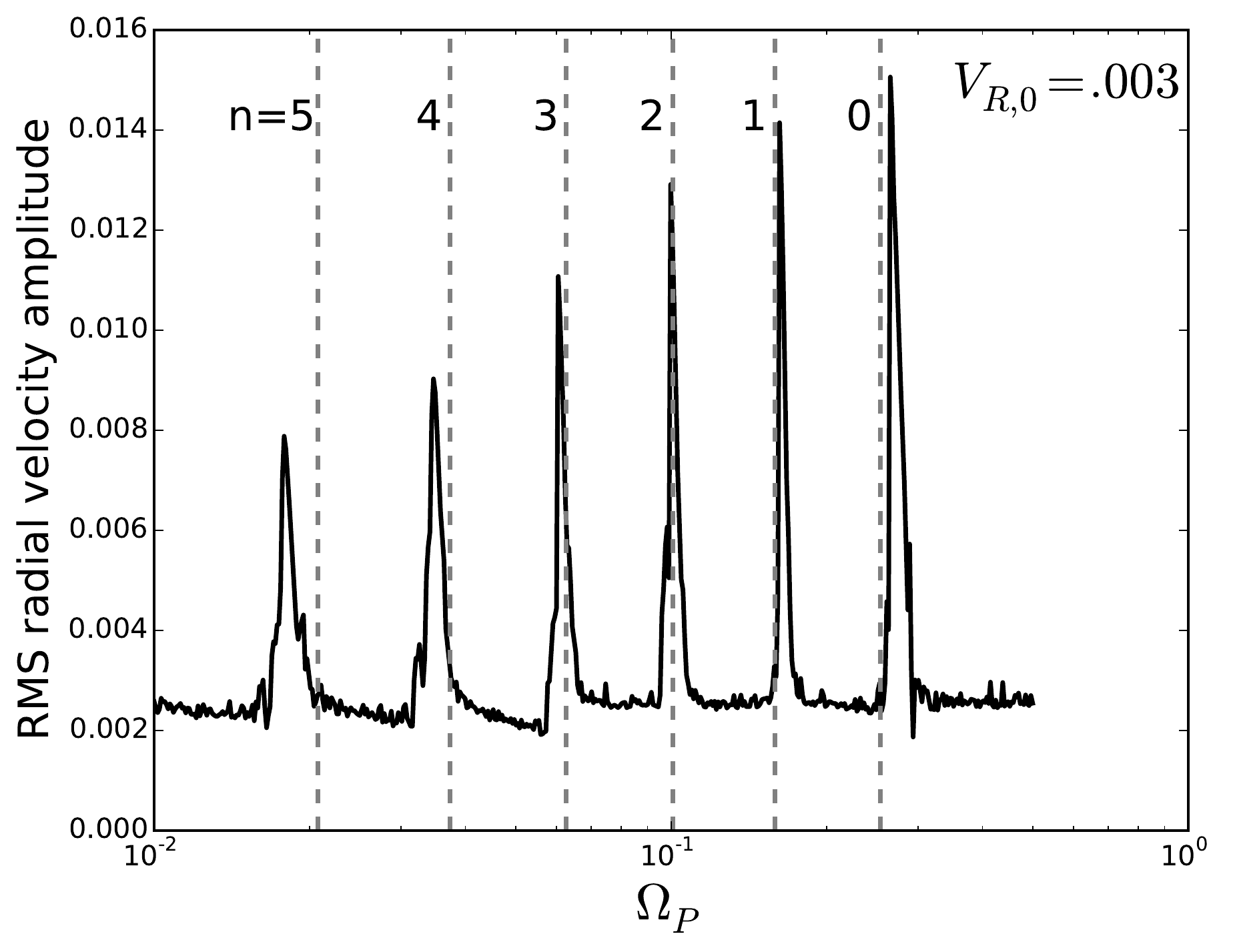}}
\caption{Root mean square average of the radial velocity between the inner boundary and the ILR as a function of $\Omega_P$. Left panel is for a forcing amplitude at the inner boundary of $V_{R,0} = .01$, and right panel is for a forcing amplitude of $V_{R,0} = .003$. Note that the y-axis scale is different in the two panels.}
\label{resonancefig}
\end{figure*}
At the end of each simulation we measure the r.m.s.\ amplitude of the perturbation to the radial velocity  between the inner boundary and the ILR. Note that the ILR varies from one simulation to the next, since its radial location in the disk depends on $\Omega_P$. Fig.\ \ref{resonancefig} shows the r.m.s.\ radial velocity amplitude as a function of $\Omega_P$. The left panel is for the suite of simulations with $V_{R,0} = .01$ and the right panel is for the suite with $V_{R,0} = .003$. The solid lines correspond to simulation results, and the gray dashed lines show the theoretical values of $\Omega_P$ for which there is constructive interference according to the resonance condition (\ref{resonance_eq}). The labels indicate the value of $n$ (number of nodes) for each resonance from $n=0$ to $n=5$. 

\begin{figure*}[!t]
\centering
\subfigure{\begin{overpic}
		 [width=.325\textwidth]{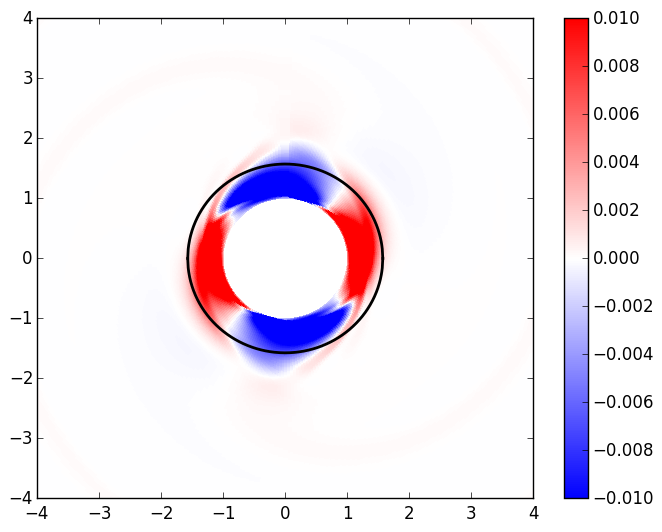}
		 \put(8,70){$n=0$}
		 \end{overpic}}
\subfigure{\begin{overpic}
		 [width=.325\textwidth]{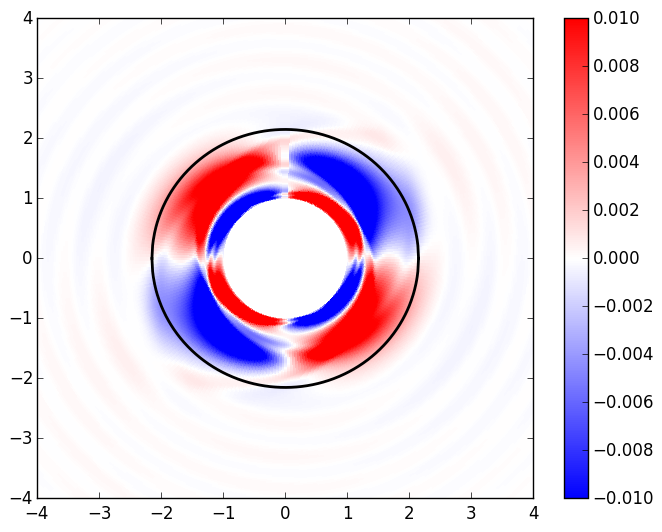}
		 \put(8,70){$n=1$}
		 \end{overpic}}
\subfigure{\begin{overpic}
		 [width=.325\textwidth]{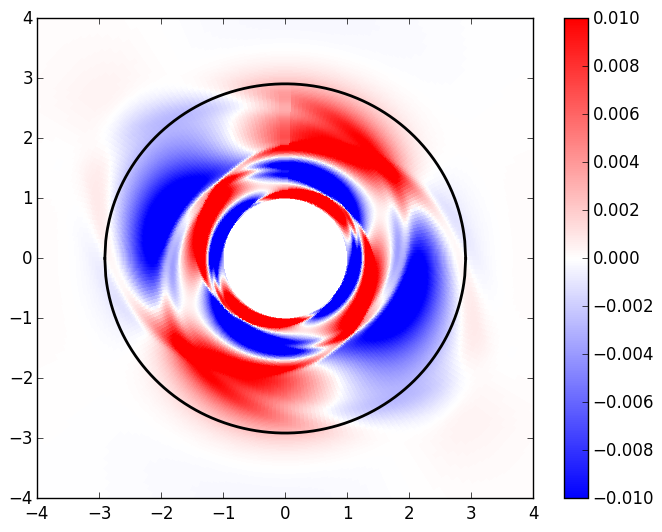}
		 \put(8,70){$n=2$}
		 \end{overpic}}
\subfigure{\begin{overpic}
		 [width=.325\textwidth]{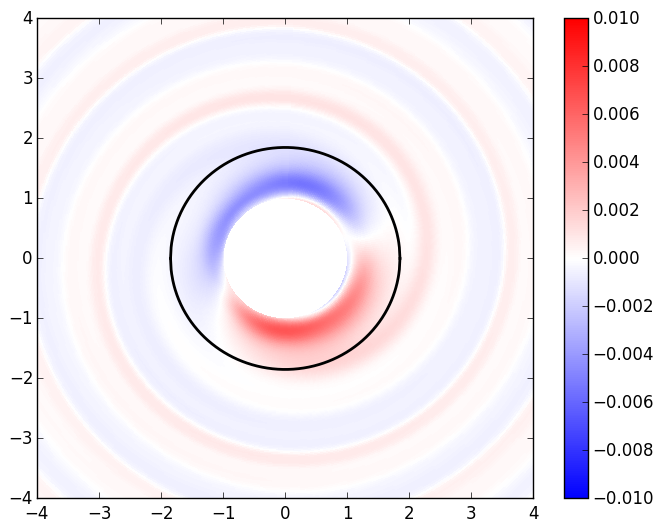}
		 \put(8,70){$n=.5$}
		 \end{overpic}}
\subfigure{\begin{overpic}
		 [width=.325\textwidth]{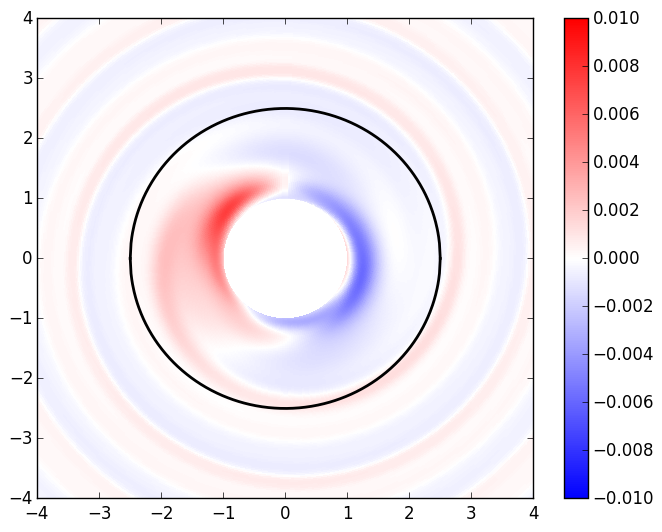}
		 \put(8,70){$n=1.5$}
		 \end{overpic}}
\subfigure{\begin{overpic}
		 [width=.325\textwidth]{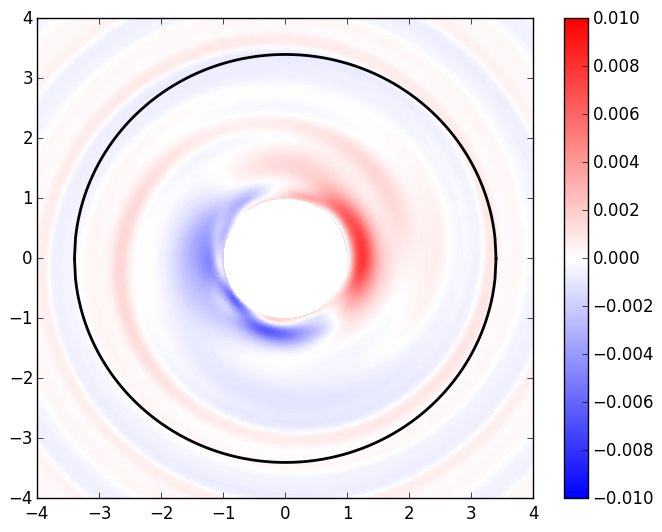}
		 \put(8,70){$n=2.5$}
		 \end{overpic}}
\caption{Radial velocity for modes with different values of $n$ from equation (\ref{resonance_eq}) at $t=3000$ for the suite of simulations with $V_{R,0} = .003$. The ILR in each panel is depicted by the black circle.}
\label{compare_fig}
\end{figure*}

In both panels of Fig.\ (\ref{resonancefig}) we see there is amplification of the r.m.s.\ amplitude of the mode near resonance by a factor of several. The theoretical predictions using the tight-winding approximation for the locations of the resonance peaks do not agree exactly with the simulation results, but do a fair job considering that the this assumption is violated near the ILR where $k_R = 0$. To visualize the structure of a trapped mode both near resonance and out of resonance, we show images of the radial velocity as a function of $(r,\phi)$ for modes with different values of $\Omega_P$ in Fig.\ \ref{compare_fig}. The modes are labeled by their value of $n$, which is implicitly defined through equation (\ref{resonance_eq}) for a given value of $\Omega_P$. Modes that are in resonance have significantly higher amplitudes than modes that are out of resonance. 

From an astrophysical point of view, the significance of the resonance condition is that it makes the problem of mode excitation {\it overdetermined}. In particular, boundary layer modes have a discrete spectrum of wavenumbers and frequencies. However, whether or not a given mode is resonant in the disk (and hence its amplitude) depends on whether or not it has an integer value of $n$ as determined by equations (\ref{resonance_eq}) and (\ref{dphi_eq}). Thus, it is possible that the relative amplitudes of boundary layer modes depend on how close they are, by chance, to having an integer value of $n$ in the disk. This is a new feature of cylindrical geometry that is not present in Cartesian geometry and comes about due to reflection of the wave from the ILR. We saw in \S \ref{star_disk_sec}, that even in the case when we do not explicitly set the pattern speed of the mode at the inner boundary, modes with integer values of $n$ are preferentially excited. In particular, the $n = 6$ slow mode that was observed in the simulation is close to resonance in the disk.

\section{Discussion}
\label{discussion}

In this work, we presented the acoustic CFS instability, which is a hydrodynamical mechanism for exciting incompressible modes in the presence of a supersonic shear layer. The acoustic CFS instability couples an incompressible mode (e.g.\ a g-mode or an r-mode) to a p-mode across the shear layer, which has a velocity drop across it that is greater than the sound speed. The shear layer could be the boundary layer between an accretion disk and a central compact object that has a material inner boundary.

We have explicitly solved for the dispersion relation in the case of a surface gravity wave (\S \ref{surfgravsec}) and a shear-Rossby edge wave (\S \ref{rossbysec}) excited by the acoustic CFS instability in a plane-parallel setup. However, the instability should generically operate for a subset of incompressible modes that radiate acoustic energy across a supersonic shear layer. As an example, consider the concrete case of an incompressible mode in a star that excites an acoustic mode in a Keplerian disk. Modes for which the pattern speed satisfies $\Omega_P > 0$ and $\Omega_* < \Omega_P < \Omega_K(R_*)$ have a positive associated energy and angular momentum density in the the star, but source an acoustic wave in the disk that carries a negative energy and angular momentum flux. Thus, the amplitude of the mode grows resulting in instability. We provide a more in depth discussion of the negative energy interpretation in Appendix \ref{negative_energy_sec}. However, one generic result is that edge modes which have their energy and angular momentum concentrated near the shear layer in the incompressible fluid are excited most efficiently by the acoustic CFS instability.

A natural application of the acoustic CFS instability is to high-frequency QPOs in neutron star and black hole systems and DNOs in non-magnetic CV systems. Both high-frequency QPOs and DNOs are associated with timescales on the order of the dynamical timescale in the innermost parts of the disk. However, DNOs are significantly more coherent and have quality factors of $\Delta \omega / \omega \sim 10^4 - 10^6$ versus $\Delta \omega / \omega \sim 10$ for QPOs. This may favor a stellar or boundary layer origin over a disk origin for DNOs. Because of the rich phenomenology associated with the oscillations in each of these astrophysical systems, we focus on DNOs in white dwarf systems. However, we first discuss how the acoustic CFS instability could apply to the case of black hole accretion.

Even though there is no apparent physical boundary in the case of black hole accretion, a shear layer should be present at the transition between a thin disk and a hot radiatively inefficient accretion flow. These two components have different pressures, and the azimuthal velocity must adjust across the transition layer to maintain hydrostatic equilibrium.  If the change in velocity satisfies $\Delta v_\phi/c_\text{s,disk} > 1$, where $c_\text{s,disk}$ is the sound speed in the cooler disk component, a mode in the hot accretion flow could be driven unstable by emission of acoustic radiation into the disk. 

This scenario seems particularly well-suited to an evanescent edge mode existing in the hot flow near the interface. In fact, there is theoretical and observational motivation for high-frequency QPOs in black hole systems being excited near the boundary between the accretion disk and the hot inner torus \citep{FragileStraubBlaes,QPO_iron_line}. Whether or not the acoustic CFS mechanism can actually excite QPOs at the disk-torus boundary could be determined in the future via global simulations of black hole accretion that include radiation transport \citep{JiangStone,Sadowski}. We mention also that \cite{LN,TsangLai} considered the related problem of excitation of high-frequency QPOs at a disk-magnetosphere interface. However, they focused on Rayleigh-Taylor and Kelvin-Helmholtz instabilities rather than destabilization of modes by emission of energy and angular momentum in the form of waves.

Switching focus to white dwarf systems, waves in the boundary layer and the disk, as well as stellar pulsations have been proposed as an explanation for DNOs \citep{PPDNO,PiroBildsten1}. Attention was paid to matching the frequencies and frequency drifts of DNOs during outburst, but the excitation mechanism of waves responsible for DNOs was typically not discussed in detail. The acoustic CFS instability provides a concrete mechanism for exciting incompressible modes in the star and in the boundary layer that could be responsible for DNOs\footnote{The stellar p-modes excited in the simulations of \cite{BRS,BRS1,BRS2} are too high frequency to explain DNOs.}. We mention that because of the generality of the acoustic CFS mechanism, it can be applied to both axisymmetric waves in a spreading layer geometry ($r-\theta$ plane) \citep{PiroBildsten1,PhilippovRafikov} and non-axisymmetric waves in a boundary layer geometry ($r-\phi$ plane) \citep{BRS1,BRS2,HertfelderKley}, as long as the flow over the surface of the star is supersonic.

Future work should focus on understanding which modes are excited in a given setup. This can help improve the connection between waves and DNOs and discriminate between competing models, such as the low-inertia magnetic accretor model \citep{Paczynski78,WarnerWoudt02} or a non-axisymmetric disk bulge \citep{Popham99}. Our work has provided a clue in this respect, though, since we showed that stellar or boundary layer modes that have a resonance in the accretion disk can be excited to substantially higher amplitudes than modes that do not. Because the spectrum of stellar modes is discrete, there may be times during a CV outburst when no stellar modes satisfy the disk resonance condition and times when several stellar modes satisfy it. This could help explain the appearance and disappearance of DNOs and the 1:2:3 frequency splitting in VW Hyi \citep{WW2005}. However, a complete picture for the origin of oscillations in non-magnetic CVs likely requires stratified 3D MHD simulations in the vein of \cite{Romanovaetal}, but with a thinner disk ($H_d/R \ll 1$) and a star that is self-consistently modeled on the grid. 

Having described the general mechanism of the acoustic CFS instability and its relevance for exciting QPOs and DNOs, we finally compare it to related astrophysical instabilities. We start with the Papaloizou-Pringle instability \citep{PP} and its plane-parallel analog \citep{Glatzel,BR}. The PPI can operate by over-reflection of trapped p-modes via tunneling through the forbidden region around corotation \citep{NGG} or by interaction with a Rossby-like mode trapped at corotation \citep{TsangLai08,TsangLai09,LaiTsang09}. \cite{LGN} found that the presence of vertical nodes in the wave has a strong stabilizing effect due to absorption of the wave at corotation. Fundamentally, though, the PPI differs from the acoustic CFS instability: the former couples two p-modes across a corotation resonance (critical layer in the plane-parallel case), whereas the latter couples a p-mode in the disk to an incompressible mode in the star.

Another related instability is the Rossby wave instability (RWI), which couples two counter-propagating Rossby-waves (with respect to the local fluid velocity) on either side of corotation \citep{RWI,RWI1,RWI2}. Since the direction of Rossby wave propagation in a barotropic fluid is determined by the gradient of the vortensity, counter-propagating Rossby waves necessary for operation of RWI exist in the vicinity of a vortensity extremum. Thus, the RWI in a rotating system can be viewed as the analog to the Kelvin-Helmholtz instability (KHI) for a shear profile of finite width in the plane-parallel case. This is because the mechanism of KHI can be interpreted as the interaction of counter-propagating shear Rossby edge-waves on either side of a critical layer \citep{Heifetz}. Again, the RWI is clearly different from the acoustic CFS instability, since the former couples two Rossby waves and requires a vortensity extremum. However, there can be some leakage of acoustic radiation past the Lindblad radii where the Rossby waves are localized \citep{RWI3}.

The acoustic CFS mechanism differs from the RWI and KHI as well as from the PPI. As its name implies, it shares the most in common with its gravitational wave counterpart \citep{CFS1,CFS2}. However, the destabilizing mechanism is the emission of acoustic waves into a fluid medium (e.g.\ an accretion disk) as opposed to emission of gravitational waves into space. We mention that \cite{HoLai} proposed an Alfv\'enic variant of the CFS instability for limiting the rotation rate of young neutron stars. In this variant, Alfv\'en waves in the magnetosphere rather than gravitational waves in vacuum or acoustic waves in a disk excite modes on a rapidly rotating neutron star.

\section*{Acknowledgements}
The author would like to thank Eliot Quataert and Roman Rafikov for stimulating discussions that helped to improve the paper. MB was supported by NASA Astrophysics Theory grant NNX14AH49G to the University of California, Berkeley and the Theoretical Astrophysics Center at UC Berkeley. This research is also funded in part by the Gordon and Betty Moore Foundation through Grant GBMF5076. This research used the SAVIO computational cluster resource provided by the Berkeley Research Computing program at the University of California, Berkeley (supported by the UC Berkeley Chancellor, Vice Chancellor of Research, and Office of the CIO). This work has benefited from the support provided by the NASA grant NNX15AR18G and NSF grant 1515763. Resources supporting this work were provided by the NASA High-End Computing (HEC) Program through the NASA Advanced Supercomputing (NAS) Division at Ames Research Center.
 
\appendix

\section{Physical Interpretation of the Growth Rate in the Hypersonic Limit}
\label{appendix_sec}
\subsection{Gravity Wave Sourced by Acoustic CFS Mechanism}
\label{hypersonic_grav}
We can understand the imaginary term in the dispersion relation (\ref{disrel4}) leading to damping or instability in the hypersonic limit using a physical argument. Consider again a normal mode perturbation. In the hypersonic limit, we can directly relate the rate of change of the energy and momentum density of the gravity wave in the lower fluid to the energy and momentum flux carried by the sound wave in the upper fluid. 

For specificity, let us consider the momentum density and the momentum flux. Using angle brackets to denote an average over one period, the growth rate is
\begin{align}
{\rm Im}[\omega] &= \frac{1}{2} \frac{{\rm d}\ln \langle \mathcal{P}_{\rm w} \rangle}{{\rm dt}} \\
&= - \frac{1}{2} \frac{\langle \mathcal{F}_P \rangle}{\langle \mathcal{P}_{\rm w} \rangle}.
\label{Imw2}
\end{align}
Here, $\mathcal{P}_{\rm w}$ represents the momentum density of the gravity wave in the lower fluid (which is in the $y$-direction) {\it per unit area} (i.e.\ integrated over $x$). Likewise, $\mathcal{F}_P$ represents the flux of $y$-momentum in the $x$-direction carried by the sound wave in the upper fluid per unit area (i.e.\ it represents the $xy$-component of the stress-tensor). The factor of $1/2$ accounts for the fact that the momentum density is a second order quantity, and since we are considering a normal mode perturbation, it has an exponential growth rate that is double that of a first order quantity.

Defining the first order fluid element displacement in the $x$-direction in the lower fluid as
\begin{align}
\label{xi_def_lower}
\delta \xi_-(x,y,t) \equiv \delta \xi_0  \cos(k_y y - \omega t)\exp(k_y x),
\end{align}
the y-momentum density in the lower fluid is
\begin{align}
\langle \mathcal{P}_{\rm w} \rangle &= \rho_- \left \langle \int_{-\infty}^{\delta \xi_-(0,y,t)} dx \delta v(x,y,t) \right \rangle, \\
&= \rho_- \left \langle \int_{-\infty}^{0} dx \delta v(x,y,t) \right \rangle + \rho_- \left \langle \delta \xi_-(0,y,t) \delta v(x,y,t) \right \rangle \\
&= \rho_- \left \langle \delta \xi_-(0,y,t) \delta v(x,y,t) \right \rangle.
\label{Pwdef}
\end{align}
The last step follows from setting the averaged mass flux equal to zero. This is true when the perturbation velocity field is irrotational and can be expressed as the gradient of a potential \citep{Phillipsbook}.

Using the condition of incompressibility, $ \bfnabla \cdot \bfv = 0$, we can write in the lower fluid
\begin{align}
\delta v = i \delta u = \omega \delta \xi_-. 
\label{incompress_xiv}
\end{align}
With this relation, equation (\ref{Pwdef}) for the gravity wave momentum density in the lower fluid is
\begin{align}
\langle \mathcal{P}_{\rm w} \rangle &= \frac{1}{2} \rho_- \omega {\delta \xi_0}^2.
\label{Pw}
\end{align}
In deriving the result (\ref{Pw}), we have used an Eulerian approach by evaluating fluid quantities at a fixed point in space in the lower fluid. An alternate approach is to use a Lagrangian formulation and evaluate the distribution of momentum flux carried by individual fluid elements \citep{Phillipsbook}. This yields the same answer as equation (\ref{Pwdef}) and describes the Stokes drift of individual fluid elements.

Turning now to the upper fluid, the momentum flux carried by the sound wave is
\begin{align}
\langle \mathcal{F}_P \rangle &= \langle \rho_+ \delta u (V_+ + \delta v) \rangle, \\
&= V_+ \langle \rho_+ \delta u\rangle + \langle{\rho_+ \delta u \delta v} \rangle \\
&= \langle{\rho_+ \delta u \delta v} \rangle
\label{Fp_incomplete}
\end{align}
In going from the second line to the third line, we have used the fact that the averaged mass flux in the sound wave is zero to second order. 

Defining the fluid element displacement in the $x$-direction in the upper fluid as 
\begin{align}
\label{xi_def_upper}
\delta \xi_+(x,y,t) \equiv \delta \xi_0  \cos(k_{x,+} x + k_y y - \ol{\omega} t),
\end{align}
we can write 
\begin{align}
\label{dudxi}
\delta u = \ol{\omega} \delta \xi_0  \sin(k_{x,+} x + k_y y - \ol{\omega} t).
\end{align}
Also, because the wave in the upper fluid is a sound wave, we have
\begin{align}
\label{dvdxi}
\delta v = \frac{k_y}{k_{x,+}} \ol{\omega} \delta \xi_0  \sin(k_{x,+} x + k_y y - \ol{\omega} t). 
\end{align}
We also need to specify the sign of $k_{x,+}$ corresponding to outgoing waves in the upper fluid. Using the boundary condition for outgoing radiation (equation (\ref{rad_out1})) together with equations (\ref{dudxi} \& \ref{dvdxi}), we can write equation (\ref{Fp_incomplete}) as
\begin{align}
\langle \mathcal{F}_P \rangle &= \frac{1}{2} \rho_+ \frac{ k_y }{k_{x,+}} \ol{\omega}^2 \delta \xi_0^2.
\label{Fp}
\end{align}

Combining equations (\ref{Imw2}), (\ref{Pw}), and (\ref{Fp}), we can write
\begin{align}
{\rm Im}[\omega] &= -\omega \left[ \epsilon \frac{1}{2} \frac{k_{x,+}}{k_y} \left( \frac{\ol{\omega}}{\omega} \right)^2 \right].
\end{align}
To further simplify the expression for ${\rm Im}[\omega]$, we use the result from equation (\ref{disrel4}) that ${\rm Re}[\omega] = {\rm sgn}(\omega)\omega_0$. Since the real part is dominant in magnitude over the imaginary part, we can write
\begin{align}
{\rm Im}[\omega] &= -{\rm sgn}(\omega)\omega_0 \left[ \epsilon \frac{  {\rm sgn}(\ol{\omega})}{2} \left(\olM^2-1 \right)^{-1/2} \left( \frac{k_yV_+}{{\rm sgn}(\omega) \omega_0} - 1\right)^2 \right],
\label{disrel5}
\end{align} 
where we have used equation (\ref{k_hyper}) to substitute for $k_{x,+}$ in terms of the wave Mach number. Comparing the imaginary part of the dispersion relation (\ref{disrel4}) with equation (\ref{disrel5}) we see that they are identical. 

As an important point, we mention that although we used the $y$-momentum density and $y$-momentum flux in equation (\ref{Imw2}) for calculating the growth rate, we could just as easily have used the energy density and energy flux. This is because for surface gravity waves, the energy and $y$-momentum densities are related via
\begin{align}
\langle \mathcal{E}_w \rangle = \frac{\omega}{k_y} \langle \mathcal{P}_w \rangle
\label{EPdensrelation}
\end{align}
\citep{Phillipsbook}. Similarly, for a sound wave in the upper fluid, the energy flux in the $x$-direction is related to the $y$-component of the momentum flux via
\begin{align}
\langle \mathcal{F}_E \rangle &= \langle \delta P \delta u \rangle + \rho_+ V_+ \langle \delta v \delta u \rangle \\ 
&= c_s \frac{\left(k_{x,+}^2 +k_y^2\right)^{1/2}}{k_y} \langle \mathcal{F}_P \rangle + V_+ \langle \mathcal{F}_P \rangle \\
&= \frac{\omega}{k_y} \langle \mathcal{F}_P \rangle.
\label{EPfluxrelation}
\end{align}
In going from the second line to the third line, we have used the fact that $\ol{\omega}/\left(k_{x,+}^2 + k_y^2\right)^{1/2} = c_s$. 

Comparing equations (\ref{EPdensrelation}) and (\ref{EPfluxrelation}), we see that using the energy density and the energy flux in equation (\ref{Imw2}) yields the same growth rate. This is not surprising in light of the fact that energy and momentum must be conserved simultaneously. The proportionality factor of $\omega/k_y$ in the ratio between the energy flux and the $y$-momentum flux is analogous to the factor of $\Omega_P$ in the ratio between the energy flux and the angular momentum flux in spiral density waves \citep{BinneyTremaine}. 

\subsection{Shear Rossby Wave Sourced by Acoustic CFS Mechanism}

We now provide a physical interpretation of the hypersonic dispersion relations (\ref{oNeq1}) and (\ref{oReq1}) for the shear Rossby wave sourced by acoustic CFS instability. Adopting the same approach as in \S \ref{hypersonic_grav}, we interpret the imaginary component of the dispersion relation, which leads to growth or damping, as due to radiation of acoustic energy that propagates away to infinity through the compressible fluid. As our starting point, we shall again adopt equation (\ref{Imw2}), which relates the growth rate to the ratio between the $y$-momentum in the edge mode in the lower fluid to the $y$-momentum flux carried by the sound wave in the upper fluid. 

We begin by computing the momentum density of the incompressible edge mode in the lower fluid. We start from the basic formula
\begin{align}
\label{Pwdef_ros}
\langle \mathcal{P}_{\rm w} \rangle &= \rho_- \left \langle \int_{x_0}^{\delta \xi_-(0,y,t)} dx \left(V_y(x)+\delta v(x,y,t) \right) - \int_{x_0}^{0} dx V_y(x) \right\rangle \\
&= \rho_- \left \langle \int_{x_0}^{\delta \xi_-(0,y,t)} dx \left(Sx+\delta v(x,y,t) \right) - \int_{x_0}^{0} dx Sx \right\rangle \\
&= \rho_- \left \langle \frac{1}{2}S \delta \xi^2_-(0,y,t) + \int_{x_0}^{\delta \xi_-(0,y,t)} dx \delta v(x,y,t)  \right\rangle,
\label{Pwdef_ros1}
\end{align}
where $V_y(x) = Sx$ is the background flow velocity in the lower fluid as per equation (\ref{shear_prof_ros}), and $\delta \xi_-$ is again given by equation (\ref{xi_def_lower}). The first term in equation (\ref{Pwdef_ros}) represents the momentum density per unit area in the presence of the mode, and the second term is the same quantity in the absence of the mode, i.e. for the equilibrium background state. Thus, their difference is the momentum density attributed to the mode. The lower integration limit $x_0 < 0$ is chosen such that $|k_y x_0| \gg 1$. We do not set $x_0 = -\infty$, however, since there is infinite momentum density contained in the background shear in the lower fluid in our setup. However, the momentum density due to the mode is finite since $\delta v \propto \exp(k_y x)$ and is negligibly small for $|k_y x_0| \gg 1$. 

To simplify equation (\ref{Pwdef_ros1}) further, we need to express $\delta v$ in terms of $\delta \xi_-$. Using the incompressibility condition, $\bfnabla \cdot \bfv = 0$, we can write
\begin{align}
\delta v = i \delta u = (\omega-k_ySx) \delta \xi_-. 
\label{incompress_xiv_ros}
\end{align}
Substituting equation (\ref{incompress_xiv_ros}) into equation (\ref{Pwdef_ros1}), the incompressible mode momentum density is
\begin{align}
\langle \mathcal{P}_{\rm w} \rangle &= \rho_- \left \langle \frac{1}{2}S \delta \xi^2_-(0,y,t) + \int_{x_0}^{\delta \xi_-(0,y,t)} dx (\omega-k_ySx) \delta \xi_-(x,y,t) \right \rangle\\
&= \rho_- \left \langle \left. \frac{1}{2}S \delta \xi^2_-(0,y,t) + \int_{x_0}^{0} dx \delta v + (\omega-k_ySx) \delta \xi^2_-(x,y,t) \right|_{x=0}\right \rangle\\
&= \rho_- \left \langle \frac{1}{2}S \delta \xi^2_-(0,y,t) +  \omega \delta \xi^2_-(0,y,t) \right \rangle.
\label{Pw_def_ros2}
\end{align}
In going to the last line, we have used the fact that the averaged mass flux due to the wave equals zero. Expression (\ref{Pw_def_ros2}) evaluates to
\begin{align}
\langle \mathcal{P}_{\rm w} \rangle &= \rho_- \left(\frac{\omega}{2} + \frac{S}{4} \right){\delta \xi_0}^2.
\label{Pw_ros}
\end{align}

Next, we turn our attention to the acoustic energy radiated in sound waves. This is simply given by equation (\ref{Fp}), just as in the case of the surface gravity wave setup. Substituting the results from equations (\ref{Fp}) and (\ref{Pw_ros}) into equation (\ref{Imw2}), we find the growth/damping rate equals
\begin{align}
{\rm Im}[\omega] &= -\frac{1}{2} \frac{\langle \mathcal{F}_P \rangle}{\langle \mathcal{P}_{\rm w} \rangle} \\
&= -\epsilon \left(\frac{k_y}{k_{x,+}}\right)\left(\frac{ \ol{\omega}^2}{2\omega + S }\right) \\
&= - {\rm sgn}(\ol{\omega}) \epsilon \left(\olM^2-1 \right)^{-1/2} \left(\frac{ \ol{\omega}^2}{2\omega + S }\right),
\label{growdamp_ros}
\end{align}
where we have used equation (\ref{k_hyper}) in the last step to substitute for $k_{x,+}$ in terms of the wave Mach number.

Substituting $\omega_N \approx \text{Re}[\omega_N] = 0$ and $\omega_R \approx \text{Re}[\omega_R] = -S$ from equations (\ref{oNeq}) and (\ref{oReq}) into equation (\ref{growdamp_ros}) in place of $\omega$, we find
\begin{align}
\label{ImwN}
{\rm Im}[\omega_N] &= - {\rm sgn}(\ol{\omega}) \epsilon \left(\olM^2-1 \right)^{-1/2} \frac{\left(k_yV_+\right)^2}{S} \\
{\rm Im}[\omega_R] &= {\rm sgn}(\ol{\omega}) \epsilon \left(\olM^2-1 \right)^{-1/2} \frac{\left(S+k_yV_+\right)^2}{S}.
\label{ImwR}
\end{align}
Comparing equations (\ref{ImwN}) and (\ref{ImwR}) with the imaginary parts of equations (\ref{oNeq1}) and (\ref{oReq1}) we see that the two are identical so our intuitive physical approach has again yielded the mathematically correct result for the imaginary part of the dispersion relation. Moreover, simply examining equation (\ref{Imw2}) reveals that there is growth when the radiated acoustic momentum flux in the upper flux is of the opposite sign as the momentum density in the lower fluid. This is the essence of the acoustic CFS mechanism.

\section{Negative Energy Interpretation}
\label{negative_energy_sec}

In a moving fluid, the energy and momentum associated with a wave depend on the choice of reference frame \citep{piercewaves,Ostrovskii}. There is a distinction between the true wave energy and momentum and the wave pseudoenergy and pseudomomentum. The latter are derived from the linearized form of the wave Lagrangian, which ignores nonlinear second order corrections to the mean flow \citep{Whitham,McIntyre}. However, the linearized wave pseudoenergy is a useful concept, and for our purposes we simply refer to it as the wave energy. 

For instance, consider a plane-parallel setup where the wavevector $k$ of the perturbation is along the direction of the flow. If the wave energy in the rest frame of the fluid is $\mathcal{E}_0 > 0$ and the frequency in the rest frame of the fluid is $\omega_0$, then the wave energy in a frame where the fluid moves with velocity $V$ is given by
\ba
\mathcal{E} = \mathcal{E}_0\left(1 + \frac{k  V}{\omega_0} \right). 
\label{Ewave_cart}
\ea
This result can be derived using physical arguments \citep{piercewaves} or using a variational principle \citep{Ostrovskii}. Equation (\ref{Ewave_cart}) shows that the wave energy density depends on the frame of reference. However, in a system with shear where $k_y \Delta V_y > \omega_0$ and $\Delta V_y$ is the magnitude of the change in the velocity, there exist regions where the mode has positive energy and regions where it has negative energy. The energy density of the wave changes sign across the critical layer where $\omega_0 + k_y V_y(x) = 0$. Exchange of energy between positive and negative energy regions existing on either side of the critical layer allows the mode to grow in amplitude, while conserving the total energy of the system.

For instance, consider the example of the Chandrasekhar-Friedman-Schutz (CFS) instability. In this case, a mode with pattern speed $0 < \Omega_P < \Omega_*$ on the surface of a star rotating with angular velocity $\Omega_*$ is unstable to emission of gravitational radiation. We can understand this instability in simple terms by considering the generalization of equation (\ref{Ewave_cart}) inside the star
\ba
\mathcal{E} = \mathcal{E}_0\left(\frac{\Omega_P}{\Omega_P - \Omega_*} \right).
\label{Ewave_star}
\ea
Here, $\Omega_P = \omega_0/m + \Omega_*$ is the pattern speed in the non-rotating frame, and $\omega_0/m$ is the pattern speed in the frame corotating with the star. The gravitational wave outside the star has positive energy, $\mathcal{E} > 0$ and hence carries a flux of positive energy away from the star. However, modes inside the star which are retrograde in the frame of the star but are drawn prograde by rotation ($0 < \Omega_P < \Omega_*$) have negative energy $\mathcal{E} < 0$. Hence, such stellar modes are sourced by emission of gravitational radiation. A similar argument for instability can be made by considering the angular momentum flux carried by the gravitational wave and the angular momentum density of the mode in the star.

For the acoustic CFS instability, the star is rotating more slowly than the adjoining accretion disk, so the physical setup is different. In this case, the energy of a wave in the star is again given by equation (\ref{Ewave_star}), but the energy of the wave in the accretion disk at the surface of the star is
\ba
\mathcal{E} = \mathcal{E}_0\left(\frac{\Omega_P}{\Omega_P - \Omega_K(R_*)} \right).
\label{Ewave_disk}
\ea
Comparing equations (\ref{Ewave_star}) and (\ref{Ewave_disk}), we see that if $\Omega_P > 0$ and $\Omega_* < \Omega_P < \Omega_K(R_*)$, the wave in the star has positive energy density $\mathcal{E} > 0$, but the wave in the disk has negative energy density $\mathcal{E} < 0$. Thus, emission of acoustic radiation with a negative energy density into the disk sources a wave with a positive energy density in the star. 

The negative energy interpretation highlights that the acoustic CFS instability should operate in a supersonic spreading layer scenario as well \citep{InogamovSunyaev,PiroBildsten}, where the modes are excited in the meridional rather than the azimuthal direction \citep{PhilippovRafikov}. Another generic result of the negative energy viewpoint is that edge modes, which are evanescent in the star away from the boundary layer, will be excited most efficiently. Such edge modes have their energy concentrated in the vicinity where the sound waves are excited. Since the growth rate of the acoustic CFS mechanism is proportional to the energy density per unit surface area for the mode in the star relative to the energy flux carried by the sound wave into the disk, these edge modes have the fastest growth rates.

\bibliography{../bibliography/NEWbib}

\end{document}